\RequirePackage{fix-cm}
\documentclass[smallextended,anonymous]{svjour3}       
\smartqed  
\usepackage{algorithmic}
\usepackage{graphicx}
\usepackage{textcomp}
\usepackage{xcolor}

\usepackage{xspace}
\usepackage{listings}
\usepackage{color}
\usepackage{multirow}
\usepackage{lscape} 
\usepackage{xparse}
\usepackage{placeins}
\usepackage{caption} 
\usepackage{enumitem}
\usepackage{array, makecell}
\usepackage{wrapfig}
\usepackage{colortbl}
\usepackage{tabularx} 
\usepackage{booktabs}
\usepackage{courier}
\usepackage{ragged2e} 
\usepackage{amsmath}
\usepackage{hyperref}
\usepackage{comment}
\usepackage{pdflscape}
\usepackage{afterpage}
\usepackage{fontawesome}
\usepackage{cite}

\definecolor{lightgray}{rgb}{.9,.9,.9}
\definecolor{darkgray}{rgb}{.4,.4,.4}
\definecolor{purple}{rgb}{0.65, 0.12, 0.82}
\lstdefinelanguage{JavaScript}{
  keywords={typeof, new, true, false, catch, function, return, null, catch, switch, var, if, in, while, do, else, case, break},
  keywordstyle=\color{blue}\bfseries,
  ndkeywords={class, export, boolean, throw, implements, import, this, setTimeout},
  ndkeywordstyle=\color{red}\bfseries,
  identifierstyle=\color{black},
  sensitive=false,
  comment=[l]{//},
  morecomment=[s]{/*}{*/},
  commentstyle=\color{purple}\ttfamily,
  stringstyle=\color{red}\ttfamily,
  morestring=[b]',
  morestring=[b]"
}

\newcommand{\header}[1]{\par\smallskip\noindent\textbf{#1.}}

\newboolean{showcomments}
\setboolean{showcomments}{true}
\ifthenelse{\boolean{showcomments}}
{
    \definecolor{myyellow}{RGB}{255, 228, 26}
    \definecolor{myblue}{RGB}{50, 50, 220}
    \newcommand{\nb}[2]{
        {\sf
            \fcolorbox{myyellow}{yellow}{\scriptsize\textbf{#1}}%
            $\blacktriangleright$%
            {\color{myblue}\fontsize{7pt}{8pt}\selectfont\textbf{#2}}%
        }%
    }
}
{
    \newcommand{\nb}[2]{}
}

\newcommand{\toolname}{\textsc{Reptory}\xspace}
\newcommand{\code}[1]{{\texttt{#1}}}

\makeatletter
\DeclareRobustCommand{\change}{%
  \@bsphack
  \leavevmode
  \@esphack
}
\DeclareRobustCommand{\stopchange}{%
  \@bsphack
  \normalcolor
  \@esphack
}
\makeatother

\usepackage{float}
\usepackage{rotating}
\usepackage{booktabs}  
\usepackage{siunitx}  
\usepackage{adjustbox}
\usepackage{pifont}
\usepackage{afterpage}
\usepackage{amsmath,amssymb}

\newcommand{\cmark}{\ding{51}}%
\newcommand{\xmark}{\ding{55}}%

\usepackage{caption}

\usepackage{picins}

\begin{document}

\title{
A Controlled Experiment of Different Code Representations for Learning-Based Program Repair
}

\titlerunning{A Controlled Experiment of Different Code Representations}        

\author{Marjane Namavar$^*$ \and
        		Noor Nashid$^*$ \and
        		Ali Mesbah
}


\institute{Marjane Namavar \at
             		The University of British Columbia, Vancouver, Canada,  \email{marjane@ece.ubc.ca} 
           	 \and
           		Noor Nashid \at
              	The University of British Columbia, Vancouver, Canada, \email{nashid@ece.ubc.ca}
           \and
           		Ali Mesbah \at
              	The University of British Columbia, Vancouver, Canada, \email{amesbah@ece.ubc.ca}
}

\date{Received: date / Accepted: date}

\def\thefootnote{*}\footnotetext{The first two authors contributed equally to this work.}\def\thefootnote{\arabic{footnote}}

\maketitle


\begin{abstract}
Training a deep learning model on source code has gained significant traction recently.
Since such models reason about vectors of numbers, source code needs to be converted to a code representation before vectorization. Numerous approaches have been proposed to represent source code, from sequences of tokens to abstract syntax trees. However, there is no systematic study to understand the effect of code representation on learning performance. Through a controlled experiment, we examine the impact of various code representations on model accuracy and usefulness in deep learning-based program repair. We train 21 different generative models that suggest fixes for name-based bugs, including 14 different homogeneous code representations, four mixed representations for the buggy and fixed code, and three different embeddings. We assess if fix suggestions produced by the model in various code representations are automatically patchable, meaning they can be transformed to a valid code that is ready to be applied to the buggy code to fix it. We also conduct a developer study to qualitatively evaluate the usefulness of inferred fixes in different code representations. Our results highlight the importance of code representation and its impact on learning and usefulness. Our findings indicate that (1) while code abstractions help the learning process, they can adversely impact the usefulness of inferred fixes from a developer's point of view; this emphasizes the need to look at the patches generated from the practitioner’s perspective, which is often neglected in the literature, (2) mixed representations can outperform homogeneous code representations, (3) bug type can affect the effectiveness of different code representations; although current techniques use a single code representation for all bug types, there is no single best code representation applicable to all bug types. 
\end{abstract}

\renewcommand{\arraystretch}{1.3}

\section{Introduction}
Automated Program Repair (APR) aims at generating patches for a given bug automatically \cite{legouesNFWTSE2012,article-bibliography}. 
Recently, there are new approaches proposed \cite{inproceedings-wild,Gupta2017DeepFixFC,Scott2019GetafixLT} that employ deep-learning (DL) techniques to \emph{learn} bug fixes from existing code corpora~\cite{article-semantic-code-repair, article-deep-reinforcement, article-jointly-learning}. 
Learning-based repair is also gaining traction in practice. For example, Facebook proposed SapFix~\cite{inproceedings-sapfix} that runs as part of their continuous integration infrastructure and can suggest bug fixes for runtime crashes; Google proposed DeepDelta~\cite{mesbah-deepDelta-fse-2019}, which automatically learns to suggest fixes for build-time compilation failures. The main advantage of a learning-based approach over more traditional techniques such as constraint-based~\cite{article-constraint-model, inproceedings-CB-DS-repair, article-AR-PHP}, heuristic-based~\cite{inproceedings-learned-heuristics}, search-based~\cite{legouesNFWTSE2012,Liu2018LSRepairLS, Mehne2018AcceleratingSP}, or dynamic~\cite{article-prophet} and static~\cite{inproceedings-dynamoth} analysis-based repair, is that it does not require domain-specific knowledge about bug patterns and fixes in a given programming language. 

To apply learning-based techniques to source code, it needs to be vectorized. Before vectorization, it is pertinent to first delineate which parts of the code and in what format will be included as input for vectorization. This is what we refer to as \emph{code representation}. As this is an emerging research area, we still lack a clear understanding of how to best represent source code for machine learning consumption.


Our work studies how different code representations along with embedding methods play a role in automated learning-based program repair. For our dataset, we focus on name-related bugs, such as swapped function arguments and wrong binary operator, as these have been extensively studied in the literature~\cite{pradel-deepbugs-oopsla-2018, argument-selection-defects-oopsla-2017,detecting-swapped-arguments-scam-2020,dl-identify-suspicious-return-saner-2020,argument-defects-icse-2016,identifier-renamings-msr-2011,argument-defects-issta-2011,arguments-defects-tse-2013}.  Depending on the usage context of a repair tool, fix suggestions could also be helpful for developers \cite{monperrus:automatic-patch-generation:icse:2014}. While some repair tools could be fully automated at runtime, there are repair-suggestion tools \cite{hartmann:suggesting-solutions:chi:2010, jeffrey:bugfix:icpc:2009, chandra:angelic-debugging:icse:11, kaleeswaran:minthint:icse14, koyuncu:ifixr:fse2019} that are built for human consumption. In this study, we consider both scenarios, i.e., some code representations could go back to the actual source code, while others could provide developers with useful debugging hints that is adequate to fix a bug. Through a controlled experiment, we train 21 models and investigate whether and to what extent changing code representations and embedding methods can affect the accuracy of the trained model in generating repair suggestions. In addition to accuracy, we analyze the perceived usefulness of fixes generated in various code representations through a user study. The contributions of our work include: 

\begin{itemize}
\item The first systematic controlled experiment to study the effects of six existing and eight newly proposed code representations, and three embeddings on the efficacy of patch generation in learning-based program repair.

\item The notion of heterogeneous abstractions for representing the buggy and fixed code (instead of a homogeneous representation for both), which is novel to the best of our knowledge for a sequence to sequence program repair task.

\item A qualitative study to assess the perceived usefulness, in terms of how developers perceive generated fixes in different code representations.

\item An evaluation on real-world bugs to verify whether the trained models are applicable to code written by developers.

\item A framework, called \toolname \cite{reptory}, which includes our code and dataset, to assist future comparative experimental studies on deep learning-based program repair. 

\end{itemize}

\lstset{
  basicstyle=\footnotesize\ttfamily,
  breaklines=true
}

\newcolumntype{M}[1]{>{\centering\arraybackslash}m{#1}}
\definecolor{lightgray}{gray}{0.9}
\renewcommand{\arraystretch}{1.5}%

\afterpage{

\begin{landscape}
\begin{table}[h]
    \leftskip=-6cm 
    
    \caption{%
    \leftskip=-0.5cm
    Different code representations for the example of Listing \ref{lst:swappedArgumentExample}
    } 
    \label{tab:code-representations}
    \begin{tabular}{M{2.4cm} M{1.3cm} m{6.8cm} m{11cm}}
    
    \toprule  
	\upshape \textbf{Representation ID (RID)} & \textbf{Category} & \textbf{Representation} & \textbf{Example}\\    
    \hline
    
WT1 & \multirow{10}{*}{\rotatebox[origin=c]{90}{\textbf{Token based}}} & Word tokenization \cite{hata-learning-to-generate-corrective-patches-2018, gupta-deepfix-aaai-2017,Hajipour2019SampleFixLT}
& \lstinline|setTimeout ( delay ,  fn )| \\ 

WT2 & & Enhanced word tokenization \cite{inproceedings-coconut,cure-program-repair-icse-2021}
& \lstinline|set <CAMEL> Timeout ( delay ,  fn )| \\ 
    
DB1 & & DeepBugs \cite{pradel-deepbugs-oopsla-2018}
& \lstinline|ID setTimeout ( ID delay , ID  fn )| \\

DB2 & & DeepBugs - arguments with types
& \lstinline|ID setTimeout ( ID number delay , ID function  fn )| \\

DB3 & & DeepBugs - arguments types only
& \lstinline|ID setTimeout ( ID number , ID function )| \\ 

FS1 & & Function signature
&\lstinline|setTimeout ( number , function  )| \\ 

FS2 & & Function signature with position anchors
&\lstinline|setTimeout ( arg0 number , arg1 function  )| \\ 

FS3 & & Function signature with LIT/ID
&\lstinline|setTimeout ( ID number , ID function  )| \\ 

FS4 & & Function signature with position anchors and LIT/ID
&\lstinline|setTimeout ( arg0 ID number , arg1 ID function  )| \\ 


TF1 & & Renaming Based Abstraction - Tufano et al. \cite{tufano:tosem:19, Learning-Meaningful-Code-Changes-Via-NMT-2019-7, inproceedings-wild}
&\lstinline|setTimeout ( Number_1 ,  Method_1  )| \\ 
\hline

AST1 & \multirow{4}{*}{\rotatebox[origin=c]{90}{\textbf{AST based}}} & AST of original code \cite{mesbah-deepDelta-fse-2019,tarlow-graph-2-diff-2019,Brody2020ASM,Ahmed2019LearningLP,deepcom,Scott2019GetafixLT}
& \lstinline|Program ExpressionStatement CallExpression Identifier setTimeout Identifier delay Identifier fn|  \\

AST2 & & AST - arguments with synthesized types
& \lstinline|Program ExpressionStatement CallExpression  Identifier setTimeout Identifier number delay Identifier function fn|  \\

AST3 & & AST - synthesized arguments types only
& \lstinline|Program ExpressionStatement CallExpression Identifier setTimeout Identifier number Identifier function|  \\

AST4 & &
Pre-order AST of original code \cite{inproceedings-6}
& \lstinline|CallExpression Identifier Identifier Identifier|  \\

    \bottomrule
    \end{tabular}

\end{table}

\end{landscape}
}

Our findings show that, as expected, a higher level of abstraction leads to better accuracy, however it can adversely affect the usefulness of the patch. The notion of mixed-representation is promising as it leads to better learning accuracy depending on the bug pattern. Additionally, our results highlight that commonly used evaluation metrics from machine learning, such as model accuracy, cannot be a sole proxy to measure the efficacy of patch generation; patchability and perceived usefulness of an inferred fix suggestion are just as important, which is largely neglected in the literature. Even though current techniques rely on a single code representation for all bug patterns, our study highlights that the code representation needs to be selected depending on the bug type as there is no single code representation that is universally effective across all bug patterns. 


\section{Motivating Example} 
\label{sec:motivation}

Listing \ref{lst:swappedArgumentExample} illustrates a real-world code example from the Angular.js project of a common name-based (aka identifier-based) bug~\cite{pradel-deepbugs-oopsla-2018, argument-selection-defects-oopsla-2017,detecting-swapped-arguments-scam-2020,dl-identify-suspicious-return-saner-2020,argument-defects-icse-2016,identifier-renamings-msr-2011,argument-defects-issta-2011,arguments-defects-tse-2013}. For the \code{setTimeout} function, the arguments must be a callback function and an integer value as the delay time, respectively. Here, the two expected arguments are passed in a wrong order in the buggy code, i.e. the first argument is a delay while the second one is a callback function. The fix is to reverse the order of these two arguments as shown in Listing \ref{lst:swappedArgumentExample}.

Researchers have proposed various ways to represent source code for deep-learning program repair. Some represent source code as a simple linear sequence of tokens~\cite{chen-sequencer-2019, Learning-Meaningful-Code-Changes-Via-NMT-2019-7, inproceedings-coconut,hata-learning-to-generate-corrective-patches-2018}. Others leverage the syntactic structure of the source code using abstract syntax trees (AST)~\cite{mesbah-deepDelta-fse-2019,Dinella2020HOPPITY}. 

\begin{lstlisting} [caption={Example of a swapped argument bug},captionpos=b,label={lst:swappedArgumentExample}]
browserSingleton.startPoller (100 ,
    function (delay, fn) {
--- 	setTimeout (delay, fn); \\bug
+++ 	setTimeout (fn, delay); \\fix  
}) ;

\end{lstlisting}


\autoref{tab:code-representations} illustrates the code statement of the example in Listing \ref{lst:swappedArgumentExample} represented in various ways. For instance, the first row depicts the representation (WT1) with simple word tokenization, whereas AST1 represents the code using the AST of the code snippet. 
In addition, the way the buggy and the fixed statements are presented can also vary. For example, we can opt for representing the buggy statement via word tokenization and the fixed statement as an AST. Considering various formats and ways of combining buggy and fixed code, there are numerous permutations of the representations imaginable for learning-based program repair. 

Given the various ways code can be represented for vectorization, in this work, we set out to explore the effect of code representations on learning repairs. Additionally, current studies on learning-based program repair have not considered the interplay between code representation and the usefulness of a fix suggestion, to the best of our knowledge. We examine whether a given code representation preserves all the fix-ingredients to generate the patch automatically. We further explore the importance of viewing patch generation from a practitioner’s perspective and ask developers to evaluate the effectiveness of a code representation to write the bug fixing commit. 

\section{Experimental Design}

The goal of our study is to assess the impact of various types of program representation on the accuracy of learning name-based bug repair, which reasons about names based on a semantic representation to suggest a fix for a piece of code that is an instance of a name-related bug pattern. We, therefore, design a controlled experiment~\cite{wohlin2012experimentation} in which we train a large set of models to compare. Using the same dataset of buggy and fixed programs as input, we change one factor of the experiment at a time, such as how the code is represented, and control for all the other variables. We then compute the accuracy of each trained model against the same test dataset, which the models have not seen before. Our study addresses the following research questions:

\afterpage{
\begin{landscape}
\begin{table}
\caption{Experimental matrix. Each row trains a new model.}

\resizebox{\linewidth}{!}{\footnotesize%
\begin{tabular}{m{0.5cm}|m{3cm}|m{7cm}|m{7cm}|m{3cm}}
\toprule

\multirow{2}{*}{\textbf{EID}} & \multirow{2}{*}{\textbf{Independent variable}}      & \multirow{2}{*}{\textbf{Abstract representation of the buggy code}}                & \multirow{2}{*}{\textbf{Abstract representation of the fixed code}}                & \multirow{2}{*}{\textbf{Embedding}}    \\
                    &                                       &                                                                           &                                                                           &                            \\ 
\midrule                    
                    
E1               & \multirow{12}{*}{\textbf{Representation}}      	& Word tokenization (WT1)                                                         					& Word tokenization  (WT1)                                                                    					& Word2Vec-CBOW     \\ \cline{1-1} \cline{3-5} 
E2              &                                                                            	& Enhanced word tokenization (WT2)                                       					& Enhanced  word tokenization (WT2)                                                  					& Word2Vec-CBOW     \\ \cline{1-1} \cline{3-5} 
E3              &                                       									& DeepBugs  (DB1)                                                   									& DeepBugs (DB1)                                                               									& Word2Vec-CBOW     \\ \cline{1-1} \cline{3-5} 
E4              &                                       									& DeepBugs - arguments with types (DB2)       							             & DeepBugs - arguments with types  (DB2)                   										& Word2Vec-CBOW    \\ \cline{1-1} \cline{3-5} 
E5              &                                       									& DeepBugs - arguments types only (DB3)   						                      & DeepBugs - arguments types only (DB3)                 	& Word2Vec-CBOW     \\ \cline{1-1} \cline{3-5} 
E6              &                                       									& Function signature  (FS1)                                      									& Function signature  (FS1)                                      & Word2Vec-CBOW     \\ \cline{1-1} \cline{3-5} 
E7              &                                       									& Function signature with position anchors (FS2)   								& Function signature with position anchors      (FS2)                   & Word2Vec-CBOW     \\ \cline{1-1} \cline{3-5} 
E8              &                                       									& Function signature with LIT/ID (FS3)                 									& Function signature with LIT/ID          (FS3)                        & Word2Vec-CBOW     \\ \cline{1-1} \cline{3-5} 
E9              &                                       									& Function signature with position anchors and LIT/ID (FS4) 					& Function signature with position anchors and LIT/ID (FS4)             & Word2Vec-CBOW    \\ \cline{1-1} \cline{3-5} 
E10            &                                       									& Renaming based abstraction (TF1)          & Renaming based abstraction (TF1)                                                         & Word2Vec-CBOW     \\ \cline{1-1} \cline{3-5}
E11            &                                       									& AST of original code (AST1)                                                   					& AST of original code      (AST1)                                                           & Word2Vec-CBOW     \\ \cline{1-1} \cline{3-5} 
E12             &                                       									& AST - arguments with synthesized types (AST2)               					& AST - arguments with synthesized types (AST2)               & Word2Vec-CBOW     \\ \cline{1-1} \cline{3-5} 
E13            &                                      									& AST - synthesized arguments types only (AST3)  				                  & AST - synthesized arguments types only (AST3)            & Word2Vec-CBOW     \\ \cline{1-1} \cline{3-5}
E14            &                                       									& Pre-order AST of original code (AST4)                                             		& Pre-order AST of original code (AST4)                                                   & Word2Vec-CBOW     \\
 \hline

E15            & \multirow{4}{*}{\textbf{Mixed representation}} & First most accurate representation $\rightarrow$ buggy code representation    & Second most accurate representation $\rightarrow$ fixed code representation                     & Word2Vec-CBOW     \\ \cline{1-1} \cline{3-5} 
E16            &                                       & Second most accurate representation $\rightarrow$ buggy code representation        							& First most accurate representation $\rightarrow$ fixed code representation                      & Word2Vec-CBOW     \\ \cline{1-1} \cline{3-5} 
E17            &                                       & Word tokenization                                                         & AST of original code                                                                       & Word2Vec-CBOW     \\ \cline{1-1} \cline{3-5} 
E18            &                                       & AST of original code                                                    & Word tokenization                                                                 & Word2Vec-CBOW     \\ \hline
E19            & \multirow{3}{*}{\textbf{Embedding}}            & \multicolumn{2}{l|}{Most accurate combination}                                                                                                                              	& Word2Vec-SG      \\ \cline{1-1} \cline{3-5} 
E20            &                                       & \multicolumn{2}{l|}{Most accurate combination}                                                                                                                              									& GloVe                     \\  \cline{1-1} \cline{3-5}
E21           &                                       & \multicolumn{2}{l|}{Most accurate combination}                                                                                                                              									& FastText                \\
\bottomrule

\end{tabular}
}
\label{tab:matrix-of-experiments}
\end{table}
\end{landscape}
}

\newlist{researchquestions}{enumerate}{1}
\setlist[researchquestions]{label*=\textbf{RQ\arabic*}}

\begin{researchquestions}[leftmargin=1.0cm]
	\item What is the influence of different code representations on learning repairs?
	
	\item Do mixed representations of buggy and fixed statements make a difference in learning repairs?
	
	\item What is the impact of using different embeddings on the learning accuracy?
	
    \item To what extent are the code representations useful to generate the correct code?
\end{researchquestions} 

\newcommand{\toolfullname}{\textsc{Representation Factory}\xspace}


To answer the research questions, we developed a framework called \toolname (short for \toolfullname) that assists with the automated name-based bug repair task in four consecutive steps: \textit{(1) Dataset Creation:} extracts correct code examples from the given corpus and generates buggy code examples using simple code transformations (see \autoref{ssec:datacollection}).
\textit{(2) Feature Extraction:} the representations illustrated in Table \ref{tab:code-representations} are used to instantiate an experimental matrix (see \autoref{sec:matrix}). The translation of each code sample into a vector is also performed in this step.
\textit{(3) Training:} trains a model for each row in our experimental matrix.
\textit{(4) Evaluation:} given a previously unseen piece of code, this step applies the trained model obtained in the previous step to generate fixes if that piece of code contains a bug.

\subsection{Dataset of Bugs and Fixes}
\label{ssec:datacollection}

To create a dataset, we start by using the code corpus provided by Raychev et al. ~\cite{veselin-learning-programs-from-noisy-data-2016} They have collected open-source JavaScript projects from GitHub, deleted duplicate files or project forks and retained only projects that parse and are not obfuscated. This process resulted in 150,000 JavaScript files with 68.6 million lines of code. We chose JavaScript as it is one of the most popular programming languages.\footnote{https://insights.stackoverflow.com/survey/2021}
 
As observed by others, the vast amount of existing code provides ample examples of likely correct code \cite{Bielik2016PHOGPM, Nguyen2015GraphBasedSL, Predictingprogrampropertiesfrombigcode}. In contrast, it is non-trivial to obtain many code examples that suffer from a particular bug pattern. One possible approach is to manually or semi-automatically search code repositories and bug trackers for examples of bugs that match a given bug pattern. However, extending this approach to thousands of samples is highly challenging but necessary for deep learning models. Instead of manually creating training data, we generate training data fully automatically from a given corpus of code. 

The notion of artificially creating likely incorrect code relates to mutation testing  \cite{jia-survey-mutation-testing} and has been applied to introduce security vulnerabilities \cite{dolan-lava-2016, pewny-evilcoder-2016}. 
To train models, we need a large set of data points of bugs and corresponding fixes. To obtain such a set, we follow the DeepBugs approach~\cite{pradel-deepbugs-oopsla-2018}, in which the existing code present in code repositories is considered as the correct version, and it is automatically mutated using targeted bug patterns to create a buggy version; it has been shown that mutants are representative of real bugs  \cite{Just2014AreMA, Andrews2005IsMA, howclose}. 
DeepBugs employs this synthetic dataset to train a model for \emph{predicting} whether previously unseen code is buggy. We use the code-level dataset and the three types of bug patterns considered in DeepBugs. 
We use the same three bug patterns, namely swapped arguments, wrong binary operator, and wrong binary operands, considered in the DeepBugs work since they are among the most prevalent name-related bugs~\cite{hanam-discovering-bug-patterns-fse-2016, articlepan}. The information gathered in this step is used in the code representation construction explained in Section~\ref{sssec:code-representation}. 
 
\header{Swapped arguments}  
\label{sssec:calls-extraction}
Swapped argument bugs happen when developers accidentally swap the function call arguments. Listing \ref{lst:swappedArgumentExample} illustrates an example of this bug pattern.
Since JavaScript is a loosely typed language and type information cannot be inferred statically, to study the impact of including type information in code representations, we also synthesize argument types. \textit{Synthesized Argument Types}. The intent is to study the impact of including type information in code representations. To this end, for each function, its arguments' types are synthesized consistently throughout the whole dataset.  We followed the same approach as in DeepBugs to extract required information \cite{pradel-deepbugs-oopsla-2018}. \autoref{table:data-swapped} exhibits information for swapped function arguments. 

\begin{table}[h]
\caption{Collected information for swapped arguments}
\centering
\label{table:data-swapped}
\begin{tabular}{m{4cm}m{8cm}}
\toprule
\textbf{Extracted} & \textit{\textbf{Description}} \\
\midrule
Callee (C) & Function name being called. \\
Base (B) & Base object if the call is a method call; else empty string. \\
 Argument 1 (\( \boldsymbol{A_{1}} \)) & First argument passed to the function. \\
 Argument 2  (\( \boldsymbol{A_{2}} \)) & Second argument passed to the function.  \\
 Arguments (A) & Array containing arguments (with DeepBugs representation). \\
 Argument Types (AT) & Array containing types of the arguments; \textit{``unknown''} if not inferred. \\
 Synthesized Argument Types (SAT) & Synthesize type information assigned to arguments. \\
\bottomrule
\end{tabular}
\end{table}

\noindent
\textit{Callee}, \textit{Base}, \textit{Argument 1}, \textit{Argument 2}, \textit{Arguments} and \textit{Argument Types} are common between this work and DeepBugs. The rest is new in our dataset. 

\begin{table}[h]
\caption{Collected information for wrong binary operator}
\centering
\label{table:data-binop}
\begin{tabular}{m{4cm} m{8cm}}
\toprule
\textbf{Extracted} & \textit{\textbf{Description}} \\
\midrule
 Left Operand (LO) & The LHS of a binary operator. \\
Right Operand (RO) & The RHS of a binary operator. \\
 Correct Operator (CO) &The correct operator. \\
 Buggy Operator (BO) &The buggy operator. \\
Operands (OPS) & An array of left and right operands with DeepBugs representation. \\
Operand Types (OT) & An array containing types of left and right operands. \\
\bottomrule
\end{tabular}
\end{table}

\header{Wrong Binary Operator} \label{sssec:wrong-binary-operator}
The wrong binary operator bug pattern encompasses accidental developer mistakes related to binary operations. To extract the desired data, we traverse the AST of each file in the code corpus and gather the information in \autoref{table:data-binop} for each binary operation. Here, \textit{Operands}, \textit{Operand Types} and \textit{Operator} are similar to what DeepBugs uses. The rest is newly extracted in our dataset. \textit{Correct Operation} and \textit{Buggy Operation} along with other extracted information will be used to represent code in various forms that is described in Section \ref{sssec:code-representation}.

\header{Wrong Binary Operands}
Wrong binary operands bug pattern pertains to the problem of swapped operands in binary operations. We followed the same approach as what DeepBugs did to extract information \cite{pradel-deepbugs-oopsla-2018}.  \autoref{table:sbo} presents information for wrong binary operands.  

\begin{table}[h]
\caption{Collected information for swapped binary operands}
\centering
\label{table:sbo}
\begin{tabular}{m{4cm} m{8cm}}
\toprule
\textbf{Extracted} & \textit{\textbf{Description}} \\
\midrule
 Left Operand (LO) & The LHS of a binary operator. \\
Right Operand (RO) & The RHS of a binary operator. \\
 Operator (O) &The correct operator. \\
Operands (OPS) & An array of left and right operands with DeepBugs representation. \\
Operand Types (OT) & An array containing types of left and right operands. \\
\bottomrule
\end{tabular}
\end{table}

\header{Data Deduplication}
To train a model and evaluate it faithfully, we need to eliminate duplicate data points~\cite{allamanis2019adverse}. We remove all duplicates from the test dataset as well as any recurrent data points between the test and training datasets. Data deduplication  is performed after the data extraction phase and before converting the data points into specific code representations. For all bug types, after data extraction explained in \ref{sssec:calls-extraction}, if all fields retrieved for two data points in a particular bug type are similar, they are considered as duplicate data points and removed. 

\subsection{Experimental Matrix}
\label{sec:matrix}
Table \ref{tab:matrix-of-experiments} shows the 21 permutations considered in our controlled experiment. Experimental runs are shown in their order of execution. Experiments employ code representations that are introduced in \autoref{tab:code-representations}. \textit{EID} refers to each experiment ID. In each row of the matrix, one independent variable is changed, while all the others are kept constant (see section \ref{ssec:independent-variables}). 
For instance, 
we start with word tokenization (WT1) in E1 and change the code representation in row E2--E14.  
Note that in the first 14 rows, the embedding is constant, and only the representation changes regardless of what embedding method is used as the goal is to compare code representations while other conditions remain the same.

At the end of experimental run E14, the two representations that yield the best results in terms of accuracy are selected. By the best code representations, we mean the best in terms of accuracy, and throughout this work, the \textit{most accurate} and the \textit{best} is used interchangeably. We mix and match these for experiments 15 and 16, where we experiment with the effects of heterogeneous (mixed) representations for the buggy and fixed code. 
As a comparison, we also include E17 and E18, which contain the mix and match of the two least abstracted representations. 
At the end of experimental run E18, the best combination of the code representations for buggy and fixed versions is selected based on the accuracy of the trained models. To assess the effects of the various embedding methods, over the next three experimental runs (E19--E21), the embedding methods are changed for the best selected code representation. 

We consider all rows (E1--E21) for the swapped arguments; however, we omit E6--E9 and E12--E13 for the wrong binary operator and operands bug types since these code abstractions do not apply to them. Next, we elaborate on the details of the experimental matrix~\autoref{tab:matrix-of-experiments}.

\subsection{Independent Variables (IV)}
\label{ssec:independent-variables}
Multiple factors affect the accuracy of a model that is meant to suggest code fixes. Below, we describe these factors as independent variables in our experiment. 
\subsubsection{Code Representation} \label{sssec:code-representation}
We can create countless ways of representing code, and it is not feasible to consider all permutations in this work; however, we can define different high-level representation categories and include examples to cover them. We partition code representations into two main classes at a high-level, namely token based and AST based (see \autoref{tab:code-representations}). The least abstracted representations of the two classes are word tokenization and AST of original code, respectively. In token based representations, a piece of code is represented with a sequence of tokens. An AST based representation incorporates structural information of the code. Both of these categories have two sub-classes where we consider a piece of code (1) without any modifications or (2) with modifications such as adding or removing information. 
In \autoref{tab:code-representations} WT1 and WT2 are based on word tokenization and represent a code string as it is. The next 3 rows (DB1--DB3) related to DeepBugs representations belong to the second subcategory where we add/remove some information to the features. In rows FS1--FS4, we simplify code by removing some information and adding some high-level extra information such as type information. TF1 refers to the code abstraction approach presented by Tufano et al.~\cite{tufano:tosem:19, Learning-Meaningful-Code-Changes-Via-NMT-2019-7, inproceedings-wild} Finally, AST1--AST4 are tree-based representations. The cited code representations in \autoref{tab:code-representations} have been proposed in the literature, while the remaining code representations are newly presented in this paper.

We consider the following different code representations in our study: 

\header{Word Tokenization} 
Deep neural networks applied to source code expect a vector of tokens as input.
Word tokenization is a token based representation and takes a piece of code without any modifications (E1 in \autoref{tab:matrix-of-experiments}) and has been used extensively in the literature~\cite{hata-learning-to-generate-corrective-patches-2018, Gupta2017DeepFixFC,Hajipour2019SampleFixLT}. Enhanced word tokenization (E2 in \autoref{tab:matrix-of-experiments}) considers underscores,
camel letters, and numbers as separators. In addition, string and number literals except for the most frequent numbers (0 and 1) are abstracted~\cite{inproceedings-coconut,cure-program-repair-icse-2021}.

\header{DeepBugs Variants} DeepBugs~\cite{pradel-deepbugs-oopsla-2018} specifies identifier tokens and literal tokens by adding \code{ID} and \code{LIT} prefixes to them, respectively. The purpose of E3 is to assess whether complementary information affects learning. Besides \code{ID} and \code{LIT}, in E4 we add variable types as well and remove variable values in E5. 

\header{Function Signature Variations} 
Function signature exhibits token based representation with modification where 
we abstract away some of the details of the source code and add some abstract information such as position anchors to assess the learning effects. 
E6 is a representation where we only take the function signature. In E7, while we remove some information and keep the function signature, we add position anchors for the arguments as well. In E8, we add abstracted identifier and literal tokens such as \code{ID} and \code{LIT} to the function signature. Finally, in E9, we add both position anchors and \code{ID} and \code{LIT} to the function signature.

\header{Renaming Based Abstraction}
The code presentation used in E10 is borrowed from Tufano et al. \cite{tufano:tosem:19, inproceedings-wild}, which applies identifier-renaming during abstraction for program-repair. The same approach was also employed to automatically learn recurring code transformation patterns \cite{Learning-Meaningful-Code-Changes-Via-NMT-2019-7}. Since NMT models are likely to be ineffective when dealing with a large vocabulary size, code tokens are abstracted by replacing actual identifiers/literals with reusable IDs. The top-300 most frequently occurring tokens are computed first. Subsequently, other identifiers and literals are replaced with reusable IDs. Each unique identifier and literal is mapped to an ID, having the form of \verb|CATEGORY_#|, where \verb|CATEGORY| represents the type of identifier or literal (i.e., \verb|METHOD|, \verb|VAR|, \verb|NUMBER|, \verb|STRING|) and \verb|#| is a numerical ID generated sequentially for each unique type of instance within that category. These IDs are used in place of identifiers and literals in the abstracted code, while the mapping between IDs and actual identifier/literal values is saved in a map M, which allows us to map back the IDs in the code. In particular, after the abstraction process, the vocabulary contains: (i) JavaScript keywords; (ii) top-300 identifiers/literals; (iii) reusable IDs.

\header{AST Variants}
AST provides a form of representation of the program structure that can assist to reason about program syntax and semantics. AST is leveraged in a variety of learning-based approaches with applications in different domains such as code summarization~\cite{sakib-summarization-msr-2020} and  repair~\cite{mesbah-deepDelta-fse-2019,Scott2019GetafixLT,Dinella2020HOPPITY}. However, AST can be represented in a variety of ways. One simple representation is the Depth First Search (DFS) traversal of the buggy and fixed statements. However, even this simple representation can have different variations, such as whether type information is embedded into the AST nodes, with or without the arguments.
E11--E14 entail AST based representations. In E11, we take the inorder traversal of AST of a piece of code without modifications. In E12, we embed type information in the AST. Since AST is more complicated than plain token sequences of code, we also include E13, where arguments are removed, and the representation is more concise and shorter. 
In E14, a code fragment is expressed as a stream of the node types that compose its AST~\cite{article-white}. For extraction, the sub-tree rooted at a specific node is selected for a given code fragment. Next, a pre-order visit of the sub-tree is performed, which for each node encountered, prints its node type.

\subsubsection{Mixed representation}
The buggy and correct statements can each be represented differently. This is what we call heterogeneous \emph{mixed representation}. For instance, the correct code can be represented using AST, whereas the buggy code can be represented using word-level tokenization. To the best of our knowledge, the notion of mixed representations is novel and has not been explored before in any sequence to sequence program repair task. For E15--E16 of \autoref{tab:matrix-of-experiments}, we mix and match best representations out of E1--E14. As a comparison, we also add E17 and E18, where only word tokenization (WT1) and AST of the original code (AST1) are involved. These cover AST based and token based representations. 

\subsubsection{Embeddings}
Our final independent variable is embedding. Machine learning requires vector representations of the data, and embeddings indicate how tokens are represented as multidimensional numerical vectors.

For E19--E21 in \autoref{tab:matrix-of-experiments}, we select the best code representation based on accuracy and only change the embedding. There are mainly two ways for embedding text; either encode each word present in the text (e.g., Word2Vec or GloVe) or encode each sentence present in the text (e.g., ELMo \cite{article-elmo} or BERT \cite{inproceedings-bert}). Encoding sentences are more useful when there is a large amount of text data and words have different meanings in different contexts. In this study, we focus on learning from short buggy statements without context, and embedding a sentence as a whole seems irrelevant here. 
We decided to keep the context out for two reasons. Firstly, context adds many variations such as the scope of the context, various abstractions for the contextual data and different ways of representing the context and we decided to investigate it in more detail in the future. Secondly, some code expressions might have a
richer context while others do not, and it makes the comparison unfair because there is this factor of context and additional information that we do not have any control over. The purpose is to show with a specific amount of data, which representation brings more value and helps the training process more.
Therefore we concentrate only on word embedding methods.  
We incorporate the following embeddings in our experiments:

\header{Word2Vec} 
Word2Vec \cite{mikolov-word2vec-2013, word2vec-iclr-2013} is one of the most popular word embedding methods where the meaning of a word can be derived from the various contexts in which this word is used. Word2Vec has been studied heavily in different source code processing tasks ~\cite{article-white, unknown-chen}. We use both architectures of Word2Vec, namely CBOW and Skip-gram (SG), in this work. 

\header{GloVe} 
Word2Vec ignores the fact that some contextual words occur more often than others and also it only takes into consideration the local context and hence failing to capture the global context. GloVe~\cite{inproceedings-glove} is an unsupervised learning algorithm for generating word embeddings by aggregating the global word-word co-occurrence matrix from a corpus~\cite{inproceedings-glove}. Although GloVe is gaining more traction in natural language processing, to the best of our knowledge, it has not been applied to source code before.


\header{FastText} 
One drawback of existing embedding methods is that they do not take into account the morphology of words and assign a distinct vector to each word. FastText~\cite{article-fasttext1} addresses this limitation by proposing a new approach based on the skipgram model. Each word is represented as a bag of character n-grams and a vector representation is dedicated to each character n-gram. Then words are represented as the sum of these representations. Wainakh et al. compare five embeddings and FastText outperforms all~\cite{wainakh-semantic-representations-of-source-code-2019}.

\subsection{Dependent Variables (DV)}
\label{ssec:DVs}
Dependent variables are the effects of changing independent variables. We measure the following dependent variables in our experiment:

\header{BLEU Score}
To measure fix suggestion quality, we use the Bilingual Evaluation Understudy (BLEU) score~\cite{Papineni2002BleuAM}. The BLEU score is a value between 0--100. It measures how close the generated output is to the expected output.
We used BLEU-4, and smooth BLEU is computed following the method outlined in \cite{orange}.  We show the BLEU score as reported by the NMT model on the best-generated prediction (top-1) in the test dataset.

\header{Position}
For models that are capable of returning multiple suggestions, such as NMT, we measure the position of the first accurate suggestion. We collect the position of the first accurate suggestion in the list of all inferred suggestions by the model, for each bug. We report average position for each code representation.

\header{Accuracy}
Accuracy is measured as the percentage of unseen test dataset that the model is able to suggest an expected output for. 
As we employ an NMT model that is capable of returning multiple suggestions, we consider it correct if any one of these suggested candidate fixes matches the expected output.
The accuracy is measured based on the given code representation for the fixed version.
For instance, consider E3 in \autoref{tab:matrix-of-experiments} where DB1 is used as the code representation to train the model; here the expected output is also in the DB1 format. 


Accuracy, compares the produced output of the model with the expected output while they are in the same code representation. It is an indicator that shows how much accurate the produced outputs are but it does not express if the produced output is automatically patchable, meaning it can be converted to the actual fixed code automatically. Considering the notion of patchability this question arises that in case that the produced output is not automatically patchable, would it be useful to developers while debugging? Another important question is how much effort is needed to convert the produced output into the actual patchable code? 
To address these issues, we have involved the following three metrics.


\header{Automatic Patchability}
Automatic Patchability is a binary metric that indicates whether a given code representation can generate a bug fixing commit automatically. It determines if the generated suggestion, in whatever code representation, can be automatically converted to a valid patch, which in fact can be applied to the buggy code. 

Code representations can encompass varied levels of information, which make them \textit{lossless} or \textit{lossy}. 
Some code representations embed additional information, such as type information, that might not be readily available in the source code. For example, DB1 and DB2 enrich the source code (WT1) with LIT/ID and type, respectively. Renaming based abstractions such as TF1 can also be transformed back to the actual code using an additional post-processing mapping steps. Such code representation transformations are \textit{lossless}, i.e., they preserve all the necessary ingredients for an automatic repair. There are also representations that are \emph{lossy}, i.e., they remove information through abstraction and might not contain the essential ingredients to fix the bug automatically. For instance, DB3 ignores arguments and only keeps type information. We use \textit{lossless} and \textit{automatically patchable} interchangeably in this work.


\header{Perceived Usefulness}
\label{sssec:patch-score}
While a lossy generated output might not be possible to convert to the actual code-level fix, it might still be useful as a debugging suggestion to developers.
Also, an aggressive code abstraction might help the learning-based approach to extract repeating patterns from the code corpora, however, the suggestions might lose their usefulness from a developer's perspective as it might not contain the required repair ingredients. Different studies \cite{Learning-Meaningful-Code-Changes-Via-NMT-2019-7, chen-sequencer-2019} have adopted a varied level of abstraction in the literature, without addressing how useful such abstractions can be in the real world. Therefore, we also assess the notion of the perceived usefulness of a generated output through a user study.

\header{Edit Distance}
\label{sec:edit-distance}
Edit distance measures the number of insertions, deletions and substitutions to transform the produced output into the actual patchable piece of code and it might indicate  the effort required to transform the generated fix suggestion into the actual valid code. 

We can divide all representations in Table \ref{tab:code-representations} into two categories: \textit{Cat1:} token based and \textit{Cat2:} AST based. The least abstracted representation in \textit{Cat1} belongs to word tokenization (WT1), while the least abstracted representation of \textit{Cat2} is AST of original code (AST1). From the developer’s perspective, a bug-fixing commit can be generated from these least abstracted representations (WT1, AST1) with the least amount of effort, and they are the closest code representations to the actual fixes. Following previous studies~\cite{ding-patching-as-translation-ase-2020,watson-learning-meaningful-assert-statements-icse-2020}, we measure edit distance (i.e., Levenshtein distance) of \textit{Cat1} and \textit{Cat2} code representations with WT1 and AST1, respectively, to measure the closeness to the actual fix.  For example, we calculate edit distance for token-based variant DB1 with WT1. On the other hand, for \textit{Cat2} variant AST3 the edit-distance calculation is performed with the least abstracted AST representation, AST1. A distance of 0 indicates that patches are identical. The higher the value, the more effort is required to obtain the actual fix in the source code. 

\subsection{Extraneous Variables (EV)}
\label{ssec:EV}
All variables, which are not defined as independent variables, but could potentially affect the results (DV) of the experiment are called extraneous variables. Such variables should be controlled where possible. In our controlled experiment, model hyper-parameters such as the number of layers, units, training steps, batch size, dropout rate, vocabulary size, and beam width are made constants after hyperparameter optimization.  The machine and environment used for all the training and inferences also remain immutable.

\subsection{Training Models}
\label{subsec:training-models}
Researchers have recently leveraged neural machine translation (NMT) to learn patterns from existing source code. For example, SequenceR~\cite{chen-sequencer-2019} and DeepDelta~\cite{mesbah-deepDelta-fse-2019} employ NMT to learn repair patterns, while Tufano et al.~\cite{Learning-Meaningful-Code-Changes-Via-NMT-2019-7} use NMT to learn code changes. NMT uses the encoder-decoder architecture with two different neural networks. The encoder network encodes the input into thought vectors. The decoder network takes this encoded data as input and decodes it to produce the desired output. Once the NMT model is trained, given previously unseen buggy code, fix suggestions can be inferred.
In our experiments, we deploy Tensorflow NMT, which uses a deep multi-layer recurrent neural network that is unidirectional and LSTM as a recurrent unit~\cite{luong17}. Attention mechanism \cite{bahdanau-attention-iclr-2015, luong-attention-mechanism-acl-2015} enables the neural network with a wider attention span to certain parts of inputs while generating the output. We enable the scaled Luong attention~\cite{luong-attention-mechanism-acl-2015} in this work. We train our models to minimise the cross-entropy loss and then back-propagate the loss to update the model parameters. The Adaptive Gradient Optimizer (Adam)~\cite{kingma:adam-optimizaiton:iclr15} is used to update the network weights of our model. We use a beam width of 25 for the inference step to generate a list of the most likely candidate sequences.
All our experimental runs are performed on an Intel(R) Xeon(R) CPU 2.50GHz machine with 62 GB RAM.
We randomly split our dataset into test and training, with 50,000 and 100,000 files, respectively. From the training set, 10\% is used as the validation set. Following the deduplication steps described in ~\ref{ssec:datacollection}, the test dataset for swapped arguments was reduced from 752,937 to 123,323 unique data points. After deduplicating 1,670,881 samples in the test dataset for wrong binary operator and wrong binary operands, the final test dataset contained 142,503 instances. Table \ref{tab:dataset-stat} provides the number of data points in the training, validation, test, and real datasets for the three bug patterns in our study. We have performed an evaluation on a real-bugs dataset explained in \autoref{sec:realworld}.
We train 21 separate models, one for each row of \autoref{tab:matrix-of-experiments} (E1--E21). 


\header{Hyperparameter optimization}
Hyperparameters can affect the performance of an NMT model, and that is why researchers optimize for their values. However, since we are conducting a controlled experiment, we need to control for these values across models trained with various code representations. To this end, we opted for tuning with the word tokenization (WT1) code representation as it is the \emph{least} abstracted. Therefore, we constructed a search space of various configurations shown in \autoref{tab:hyperparameter-tuning}. We selected values for each hyperparameter by examining previous studies \cite{code-review-icse-2021, recommendations-for-tuning-bengio-2012}. We trained 50 different models with WT1 to determine the optimal values. Following that, we trained 10 models with the tuned set and compared them with multiple trained models with random values for the remaining code representations. The tuned set outperformed the random sets in all cases. Therefore, we employed the tuned values from \autoref{tab:hyperparameter-tuning} as an approximation for the optimal set of hyperparameter values to conduct our controlled experiment. 
With the optimal configuration for WT1, we found little improvement in the BLEU score after 25K steps. Although more abstracted representations may reach their best BLEU scores earlier than 25K steps, as this is a controlled study, we decided to keep the same stopping criterion for all models. Thus, the stopping criterion for training is the number of training steps that is fixed at 25k. The training time varies from 41.32 hours (E11) for the wrong binary operator bug type to 4.75 hours (E14) for the swapped arguments bug type. The average training time for swapped arguments, wrong binary operator, and wrong binary operands is 21.97, 18.05, and 17.55 hours.

\renewcommand{\arraystretch}{0.7}
\begin{table}
\caption{Hyperparameters and the best configuration}
\label{tab:hyperparameter-tuning}
\vspace{-0.5em}
\resizebox{\linewidth}{!}{%
\scriptsize
\begin{tabular}{llc}
\toprule
\textbf{Hyperparameter} 					& \textbf{Possible Values}			             & \textbf{Tuned Value} \\
\midrule
Embedding size 								& [128, 256, 512]									& 512   \\
Encoder layers 									& [2, 3, 4]												& 2       \\
Decoder layers 								    & [2, 3, 4]												& 2        \\
Number of units 								& [128, 256, 512]  									& 512    \\
Learning rate 									& [0.0, 0.001, 0.01, 0.1, 0.2, 1.0] 				& 0.001 \\
Dropout 											& [0.0, 0.1, 0.2, 0.4, 0.5] 							& 0.1     \\
Vocabulary size 								& [5K, 10K, 20K, 30K]  							& 30K	\\
\bottomrule
\end{tabular}
}
\vspace{-1.5em}
\end{table}

\section{Patchability and Usefulness Study}
\label{sec:usefulness-study}
The ultimate goal of an automated program repair system is to assist developers with bug fixing tasks. As shown in \autoref{tab:code-representations}, learning-based techniques could employ a plethora of lossless and lossy code representations to represent the buggy and the fixed code. While some lossy representations can help extract the recurring pattern, and DL models can learn recurring source code patterns, they could not automatically produce the fixed code. To address RQ4, we examine whether a suggested fix is (1) automatically patachable, and (2) useful from developers' perspective for debugging. Fundamentally, when the code is abstracted, it helps the model to extract fix patterns, whereas the developer might find the generated fix as not useful. An end-user might tolerate abstraction to some extent while receiving enough cues.  The goal of the user study is to find out whether the abstracted representations provide enough hints to the developer to write the final patch to be useful. To understand this notion of usefulness from a developer’s perspective, we conducted a survey.

The patchability and user study are available online as part of our replication package~\toolname \cite{reptory}.


\subsection{Bug Selection}
\label{ssec:bug-selection}
Our selection of bugs is based on the following factors: 
(1) Bugs from the same type 
can differ in their complexity and might need a different amount of effort to be fixed. To keep the comparison of different representations equitable, we present the same bug in different representations.
(2) It is not feasible to consider all bugs from our dataset. Thus we have to limit the number of samples from each bug type. 

The authors have done the automatic patchability study while developers answered the questions of the user study. With the above design considerations in mind, we have created two pools of samples for each study, and we explain relevant details in the following two subsections.

\subsection{Automatic Patchability Study}
To assess the patchability of fix suggestions, the authors manually investigated the random samples from each code representation to understand whether they are automatically patchable, meaning the fix suggestions could be transformed to the actual source code in an automated way. We have 14 different code representations for swapped arguments, 8 code representations for the wrong binary operator, and 8 code representations for wrong binary operands, which leads to a total of 30 code representations from three bug types. We have selected five random bugs from each code representation, leading to a pool of 150 ($5*30$) questions. Each question shows the buggy code and actual correct code, and fix suggestions in a given code representation. The authors examine whether the actual fixed code could be generated from the fix suggestion. For example, from token-based code representation DB1, ID/LIT could be removed to transform the code into the original patchable code. DB3 only keeps type information that could not be automatically transformed back to the actual code. For the AST-based representation, type information in AST2 could be removed from its representation, and the resulting AST could be translated back to the original code. However, for AST3, arguments are not preserved and cannot be transformed automatically. Each of the authors assessed and categorized the fix suggestions independently to understand whether fix suggestions could be converted back to the actual fixed code. If the scoring differed between authors, they discussed it and reached a consensus on the automatic patchability of a code representation.


\subsection{User Study}
Despite some of the representations being lossy, they could still be helpful. The main purpose of the user study is to assess the usefulness of lossy code representations from the developer's point of view because there is no automated way or metric to investigate that. Therefore, we measure the perceived usefulness of each of the lossy code representations, individually.
On the other side, all lossless (automatically patchable) code representations can be converted to the actual fixed code which is similar to the word tokenization code representation. Hence, we only involve the samples from the word tokenization (WT1) code representation as a representative of all lossless code representations.

We have 7, 2 and 2 lossy code representations for the swapped arguments, wrong binary operator and operands, respectively. Also, we have word tokenization for each bug type.
Therefore, there are 8, 3 and 3 code representations involved in the user study for the swapped arguments, wrong binary operator and operands, respectively. In total, 14 ($8+3+3$) code representations are involved. For each code representation, 5 samples have been randomly selected. Finally, we have 70 ($5*14$) samples in a pool of samples used for the user study. 

We selected 131 top JavaScript repositories that were also used by previous work \cite{hanam-discovering-bug-patterns-fse-2016}. These repositories are a combination of most depended-upon npm packages, most starred-by users npm packages and most starred projects. We chose contributors with equal to or more than 10 contributions for these top projects. Committing to a project's default branch, posting an issue, and proposing a pull request are all examples of contributions. There could be multiple collaborators for a GitHub repository. If a contributor's email address was available in their GitHub account, we contacted them by email. We invited 405 people to participate and they were not given any benefits. We provided guidelines on (a) how to understand the survey questions, (b) how to score a patch. 44 developers completed the study. On average, our participants had 6.8 years of general software development and 3.4 years of JavaScript experience.

Every participant received 10 randomly selected questions from our pool of questions. In each question, a buggy code, the corresponding developer fix, and a fix suggestion produced by \toolname were shown. Survey participants were asked to give a score between 1--5 (Likert-type) to indicate to what extent a fix suggestion is helpful to convert the buggy code into the actual fixed code. As we are offering a fix suggestion system and what it produces is a fix suggestion that acts as a hint for developers, they need to evaluate how difficult it is to go from the buggy version to the fixed version using the fix suggestion. 1 means not useful at all, and 5 means absolutely useful. There was an open-ended question at the end of the survey to receive additional comments about fix suggestions. From the completed responses, we collected perceived usefulness values from the participants and averaged them for each representation of each bug type to achieve the perceived usefulness score of a given code representation.

\section{Results}
\label{sec:results}
Tables \ref{tab:results-swapped-arguments}--\ref{tab:results-binary-operand} present our results for the three different bug patterns. The tables are colour-coded, and in each column, the darker the colour, the better the result for that dependent variable. For example, higher accuracies and lower positions are darker.

\begin{table}[]
\caption{Statistics for the three bug types}
\label{tab:dataset-stat}
\resizebox{\linewidth}{!}{
{\scriptsize
\begin{tabular}{llllc}

\toprule
\textbf{Bug Type}         &  \textbf{Train}      &   \textbf{Val}  & \textbf{Test} & \textbf{Real}     \\ 
\midrule           
Swapped Arguments          &   1,343,669  &   149,297     & 123,323 & 23\\ 
Wrong Binary Operator      &   3,051,474  &   339,053     & 142,503 & 37\\ 
Wrong Binary Operands 	&   3,051,474  &   339,053     & 142,503 & 35\\ 
\bottomrule
\end{tabular}
}}
\end{table}


\renewcommand{\arraystretch}{1.2}
\begin{table}[!ht]
    \caption{Results - Swapped Arguments}
    \label{tab:results-swapped-arguments}

\resizebox{\linewidth}{!}{    
    \begin{tabular}{cccccccc}
    \toprule
	\upshape &{\rotatebox[origin=c]{90}{\textbf{EID(SA)}}}
	& \rotatebox[origin=c]{90}{\textbf{RID}}
	& \rotatebox[origin=c]{90}{\textbf{Accuracy(\%)}}  
	&  {\rotatebox[origin=c]{90}{\textbf{BLEU}}}      
	& {\rotatebox[origin=c]{90}{\textbf{Position}}} \\[15pt] 
	\midrule
                                                                                           & 1                          & WT1                          & \cellcolor[HTML]{E7CBD8}20.367  & \cellcolor[HTML]{FFF2CC}55.6  & \cellcolor[HTML]{457F2C}1.4 \\
                                                                                           & 2                          & WT2                          & \cellcolor[HTML]{D7ACC0}40.078  & \cellcolor[HTML]{FCE7A8}66.2  & \cellcolor[HTML]{3E7A24}1.2 \\
                                                                                           & 3                          & DB1                          & \cellcolor[HTML]{E7CAD7}20.867  & \cellcolor[HTML]{FBE39C}69.6  & \cellcolor[HTML]{437D29}1.3 \\
                                                                                           & 4                          & DB2                          & \cellcolor[HTML]{E7CAD7}20.843  & \cellcolor[HTML]{FAE092}72.4  & \cellcolor[HTML]{447E2A}1.3 \\
                                                                                           & 5                          & DB3                          & \cellcolor[HTML]{D4A5BB}44.271  & \cellcolor[HTML]{FBE5A0}68.3  & \cellcolor[HTML]{CEE2C7}5.3 \\
                                                                                           & 6                          & FS1                          & \cellcolor[HTML]{CF9BB4}50.617  & \cellcolor[HTML]{FCE6A5}67.0  & \cellcolor[HTML]{76A363}2.8 \\
                                                                                           & 7                          & FS2 (2nd most accurate)                          & \cellcolor[HTML]{C68BA8}60.912  & \cellcolor[HTML]{F8D97C}78.8  & \cellcolor[HTML]{669751}2.3 \\
                                                                                           & 8                          & FS3                          & \cellcolor[HTML]{D09EB6}48.872  & \cellcolor[HTML]{FBE49D}69.3  & \cellcolor[HTML]{D9EAD3}5.6 \\
                                                                                           & 9                          & FS4                          & \cellcolor[HTML]{E1BECE}28.534  & \cellcolor[HTML]{FBE39B}69.9  & \cellcolor[HTML]{7BA669}2.9 \\
                                                                                           & 10                         & TF1 (1st most accurate)                          & \cellcolor[HTML]{A74E7A}99.805  & \cellcolor[HTML]{F2C333}99.9  & \cellcolor[HTML]{38761D}1.0 \\
                                                                                           & 11                         & AST1                         & \cellcolor[HTML]{EAD1DC}16.405  & \cellcolor[HTML]{FBE49F}68.6  & \cellcolor[HTML]{598D42}2.0 \\
                                                                                           & 12                         & AST2                         & \cellcolor[HTML]{E9CEDA}18.833  & \cellcolor[HTML]{FADF8D}73.9  & \cellcolor[HTML]{4B8332}1.6 \\
                                                                                           & 13                         & AST3                         & \cellcolor[HTML]{C990AC}57.706  & \cellcolor[HTML]{F6D266}85.2  & \cellcolor[HTML]{447E2A}1.3 \\
\multirow{-14}{*}{\textbf{Representation}}                                                 & \cellcolor[HTML]{B7B7B7}14 & \cellcolor[HTML]{B7B7B7}AST4 & \cellcolor[HTML]{A64D79}100.000 & \cellcolor[HTML]{F1C232}100.0 & \cellcolor[HTML]{38761D}1.0 \\ \hline
                                                                                           & 15                         & TF1 $\rightarrow$ FS2        & \cellcolor[HTML]{D2A2B9}46.666  & \cellcolor[HTML]{F9DC84}76.5  & \cellcolor[HTML]{598E43}2.0 \\
                                                                                           & 16                         & FS2 $\rightarrow$ TF1        & \cellcolor[HTML]{BE7C9C}70.725  & \cellcolor[HTML]{F9DD86}75.9  & \cellcolor[HTML]{79A567}2.9 \\
                                                                                           & 17                         & WT1 $\rightarrow$ AST1       & \cellcolor[HTML]{E8CCD8}19.913  & \cellcolor[HTML]{FBE39A}70.3  & \cellcolor[HTML]{4C8433}1.6 \\
\multirow{-4}{*}{\textbf{\begin{tabular}[c]{@{}c@{}}Mixed \\ Representation\end{tabular}}} & 18                         & AST1 $\rightarrow$ WT1       & \cellcolor[HTML]{E7CBD7}20.690  & \cellcolor[HTML]{FFF2CB}56.1  & \cellcolor[HTML]{437D29}1.3 \\ \hline
                                                                                           & 19                         & TF1 (Word2Vec-SG)            & \cellcolor[HTML]{A74E7A}99.822  & \cellcolor[HTML]{F2C333}99.9  & \cellcolor[HTML]{38761D}1.0 \\
                                                                                           & 20                         & TF1 (GloVe)                  & \cellcolor[HTML]{A74E7A}99.964  & \cellcolor[HTML]{F1C232}100.0 & \cellcolor[HTML]{38761D}1.0 \\
\multirow{-3}{*}{\textbf{Embedding}}                                                       & 21                         & TF1 (FastText)               & \cellcolor[HTML]{A74E7A}99.825  & \cellcolor[HTML]{F2C333}99.9  & \cellcolor[HTML]{38761D}1.0\\
\bottomrule
\end{tabular}
}
\end{table}

\renewcommand{\arraystretch}{1}
\begin{table}[!ht]
    \caption{Results - Wrong Binary Operator 
    }
    \label{tab:results-binary-operator}
\resizebox{\linewidth}{!}{        
    \begin{tabular}{cccccccc}
    \toprule  
	\upshape &{\rotatebox[origin=c]{90}{\textbf{EID(Oprt)}}}
	& \rotatebox[origin=c]{90}{\textbf{RID}}
	& \rotatebox[origin=c]{90}{\textbf{Accuracy(\%)}}  
	&  {\rotatebox[origin=c]{90}{\textbf{BLEU}}}  
	& {\rotatebox[origin=c]{90}{\textbf{Position}}} \\[15pt] 
	\midrule
                                                                                           & 1                          & WT1                          & \cellcolor[HTML]{E4C5D3}19.270  & \cellcolor[HTML]{FEEDBC}18.1  & \cellcolor[HTML]{99BC8B}3.5 \\
                                                                                           & 2                          & WT2                          & \cellcolor[HTML]{DDB7C9}28.579  & \cellcolor[HTML]{FFF1C8}11.3  & \cellcolor[HTML]{92B783}3.3 \\
                                                                                           & 3                          & DB1                          & \cellcolor[HTML]{E3C3D2}20.175  & \cellcolor[HTML]{FCE7A8}30.0  & \cellcolor[HTML]{8BB27B}3.2 \\
                                                                                           & 4                          & DB2                          & \cellcolor[HTML]{E3C3D2}20.186  & \cellcolor[HTML]{FAE194}42.3  & \cellcolor[HTML]{8CB27C}3.2 \\
                                                                                           & 5                          & DB3 (1st most accurate)                          & \cellcolor[HTML]{B16289}85.978  & \cellcolor[HTML]{F9DD86}50.1  & \cellcolor[HTML]{8EB47E}3.2 \\
                                                                                           & 10                         & TF1 (2nd most accurate)                          & \cellcolor[HTML]{B16389}85.787  & \cellcolor[HTML]{F2C53A}95.4  & \cellcolor[HTML]{7AA567}2.7 \\
                                                                                           & 11                         & AST1                         & \cellcolor[HTML]{E6C9D6}16.570  & \cellcolor[HTML]{F8D776}60.0  & \cellcolor[HTML]{9EBF91}3.7 \\
\multirow{-8}{*}{\textbf{Representation}}                                                  & \cellcolor[HTML]{B7B7B7}14 & \cellcolor[HTML]{B7B7B7}AST4 & \cellcolor[HTML]{A64D79}100.000 & \cellcolor[HTML]{F1C232}100.0 & \cellcolor[HTML]{38761D}1.0 \\ \hline
                                                                                           & 15                         & DB3 $\rightarrow$ TF1        & \cellcolor[HTML]{CB95AF}51.513  & \cellcolor[HTML]{F6D469}67.4  & \cellcolor[HTML]{D9EAD3}5.2 \\
                                                                                           & 16                         & TF1 $\rightarrow$ DB3        & \cellcolor[HTML]{B16289}85.922  & \cellcolor[HTML]{F8D879}58.1  & \cellcolor[HTML]{7EA86C}2.8 \\
                                                                                           & 17                         & WT1 $\rightarrow$ AST1       & \cellcolor[HTML]{EAD1DC}10.667  & \cellcolor[HTML]{F9DB83}52.4  & \cellcolor[HTML]{A6C599}3.9 \\
\multirow{-4}{*}{\textbf{\begin{tabular}[c]{@{}c@{}}Mixed \\ Representation\end{tabular}}} & 18                         & AST1 $\rightarrow$ WT1       & \cellcolor[HTML]{E5C7D5}17.659  & \cellcolor[HTML]{FFF2CC}8.4   & \cellcolor[HTML]{9ABC8C}3.6 \\ \hline
                                                                                           & 19                         & DB3 (Word2Vec-SG)            & \cellcolor[HTML]{B16289}85.983  & \cellcolor[HTML]{F9DC86}50.4  & \cellcolor[HTML]{90B580}3.3 \\
                                                                                           & 20                         & DB3 (GloVe)                  & \cellcolor[HTML]{A74E7A}99.977  & \cellcolor[HTML]{F9DC85}50.9  & \cellcolor[HTML]{9DBF8F}3.6 \\
\multirow{-3}{*}{\textbf{Embedding}}                                                       & 21                         & DB3 (FastText)               & \cellcolor[HTML]{B16289}85.988  & \cellcolor[HTML]{F9DC86}50.4  & \cellcolor[HTML]{8CB27C}3.2 \\ 
\bottomrule
\end{tabular}
}
\end{table}

\textit{EID} in these tables refers to the experimental ID. For instance, \textit{EID(SA)} of 1 in \autoref{tab:results-swapped-arguments} presents our results of the experiment for swapped arguments with \textit{EID} as 1 from \autoref{tab:matrix-of-experiments}. The \textit{RID} column indicates which code representations from \autoref{tab:code-representations} are involved for that experimental run. For experiments with mixed representations, both representations for the buggy and fixed statements are presented. 

\renewcommand{\arraystretch}{1}
\begin{table*}[]
    \caption{Results - Wrong Binary Operands}
    \label{tab:results-binary-operand}
\resizebox{\linewidth}{!}{    
\begin{tabular}{cccccccc}
\toprule  
	\upshape &{\rotatebox[origin=c]{90}{\textbf{EID(Opnd)}}}
	& \rotatebox[origin=c]{90}{\textbf{RID}}
	& \rotatebox[origin=c]{90}{\textbf{Accuracy(\%)}}  
	&  {\rotatebox[origin=c]{90}{\textbf{BLEU}}}  
	& {\rotatebox[origin=c]{90}{\textbf{Position}}} \\[20pt] 
	\midrule
                                                                                               & 1                          & WT1                          & \cellcolor[HTML]{E4C4D2}20.474  & \cellcolor[HTML]{FEEEBD}28.4  & \cellcolor[HTML]{548A3C}1.4 \\
                                                                                           & 2                          & WT2                          & \cellcolor[HTML]{DCB5C7}30.625  & \cellcolor[HTML]{FFF2CC}20.6  & \cellcolor[HTML]{81AA6F}2.1 \\
                                                                                           & 3                          & DB1                          & \cellcolor[HTML]{E3C3D2}20.899  & \cellcolor[HTML]{FBE39B}46.2  & \cellcolor[HTML]{508738}1.4 \\
                                                                                           & 4                          & DB2                          & \cellcolor[HTML]{E3C3D2}20.912  & \cellcolor[HTML]{F9DC84}58.1  & \cellcolor[HTML]{4F8637}1.4 \\
                                                                                           & 5                          & DB3 (1st most accurate)                          & \cellcolor[HTML]{B16289}85.993  & \cellcolor[HTML]{F3C845}90.4  & \cellcolor[HTML]{38761D}1.0 \\
                                                                                           & 10                         & TF1 (2nd most accurate)                          & \cellcolor[HTML]{B16389}85.817  & \cellcolor[HTML]{F2C53C}95.1  & \cellcolor[HTML]{38761D}1.0 \\
                                                                                           & 11                         & AST1                         & \cellcolor[HTML]{E6C8D5}17.789  & \cellcolor[HTML]{F6D469}71.8  & \cellcolor[HTML]{6C9B57}1.8 \\
\multirow{-8}{*}{\textbf{Representation}}                                                  & \cellcolor[HTML]{B7B7B7}14 & \cellcolor[HTML]{B7B7B7}AST4 & \cellcolor[HTML]{A64D79}100.000 & \cellcolor[HTML]{F1C232}100.0 & \cellcolor[HTML]{38761D}1.0 \\ \hline
                                                                                           & 15                         & DB3 $\rightarrow$ TF1        & \cellcolor[HTML]{C487A4}61.582  & \cellcolor[HTML]{F4CD53}83.1  & \cellcolor[HTML]{D9EAD3}3.4 \\
                                                                                           & 16                         &  TF1 $\rightarrow$ DB3        & \cellcolor[HTML]{B16289}85.995  & \cellcolor[HTML]{F3C947}89.5  & \cellcolor[HTML]{39761E}1.0 \\
                                                                                           & 17                         & WT1 $\rightarrow$ AST1       & \cellcolor[HTML]{EAD1DC}11.363  & \cellcolor[HTML]{F8D879}63.9  & \cellcolor[HTML]{5D9147}1.6 \\
\multirow{-4}{*}{\textbf{\begin{tabular}[c]{@{}c@{}}Mixed \\ Representation\end{tabular}}} & 18                         & AST1 $\rightarrow$ WT1       & \cellcolor[HTML]{E3C4D2}20.673  & \cellcolor[HTML]{FEEEBF}27.6  & \cellcolor[HTML]{52883A}1.4 \\ \hline
                                                                                           & 19                         & DB3 (Word2Vec-SG)            & \cellcolor[HTML]{B16289}85.993  & \cellcolor[HTML]{F3C947}89.5  & \cellcolor[HTML]{39761E}1.0 \\
                                                                                           & 20                         & DB3 (GloVe)                  & \cellcolor[HTML]{A74E7A}99.996  & \cellcolor[HTML]{F2C335}98.9  & \cellcolor[HTML]{39761E}1.0 \\
\multirow{-3}{*}{\textbf{Embedding}}                                                       & 21                         & DB3 (FastText)               & \cellcolor[HTML]{B16289}85.992  & \cellcolor[HTML]{F3C947}89.4  & \cellcolor[HTML]{39761E}1.0 \\

 \bottomrule
\end{tabular}
}
\end{table*}

\subsection{Code Representations (RQ1)}	
RQ1 pertains to the influence of different code representations, and we consider E1--E14 here. As the function signature-related representations do not apply to the wrong binary operator and operands bug patterns, several rows are omitted in \autoref{tab:results-binary-operator} and \autoref{tab:results-binary-operand}.

\header{Perceived usefulness of AST4 representation} In Tables \ref{tab:results-swapped-arguments}--\ref{tab:results-binary-operand}, E14 with AST4 representation has the highest accuracy of 100 percent. However, in terms of edit distance and perceived usefulness, it is significantly worse than the second-worst outcome for all three categories of bugs. For swapped arguments, AST4 achieves the lowest perceived usefulness score (1.0) in \autoref{tab:results-usefulness}. The developer survey deemed this representation to be ineffective for all bug categories and it is not automatically patchable as well. 
Furthermore, developers suggested in the survey comments that this code representation does not convey meaningful information. In \autoref{tab:results-usefulness} edit distance reaches its worst value of 88.1 in swapped arguments and 70.7 in wrong binary operator and operands, where the second-worst numbers are by far smaller. Therefore, we contemplate E14 (shown in grey colour) as an outlier and do not consider it to select the first and second-best representations in terms of accuracy. 

\subsubsection{Swapped Arguments}
Considering the first 14 rows in \autoref{tab:results-swapped-arguments}, and ignoring AST4, the first and second highest accuracies reached are 99.805\% and 60.912\% for experiments E10 and E7, which used Tufano et al. \cite{tufano:tosem:19, Learning-Meaningful-Code-Changes-Via-NMT-2019-7, inproceedings-wild} (TF1) and function signature with position anchors (FS2) as code representations, respectively. Enhanced word tokenization in E2 leads to a better accuracy compared to E1. Although the accuracy in DeepBugs representations with actual values (E3--E4) was close to E1, where word tokenization was used, there was a jump in accuracy once function arguments were abstracted away in E5. Also, function signatures in E6--E8 improved accuracy. In E11 and E12, using AST did not help much with accuracy, but interestingly accuracy increased sharply in E13 with the combination of AST and abstraction. The highest BLEU score among the first 14 experiments (E1--E14) was 99.9 in E10.

The best two positions for the swapped arguments belong to E10 and E2 using Tufano et al. \cite{tufano:tosem:19, Learning-Meaningful-Code-Changes-Via-NMT-2019-7, inproceedings-wild}(TF1) and enhanced word tokenization (WT2) code representations, respectively.

\subsubsection{Wrong Binary Operator and Operands}
As shown in \autoref{tab:results-binary-operator} and \autoref{tab:results-binary-operand} in E1--E14, the first and second top accuracies for both wrong binary operator and operands are yielded by experiments E5 and E10. Analogous to swapped arguments, enhanced word tokenization improves accuracy. In these two categories of bugs, on the one hand, accuracy experienced minor improvement as we enriched word tokenization (WT1) with further information in variations of DeepBugs representations (E3--E4). On the other hand, the best accuracy was achieved once we removed arguments in E5 with DeepBugs representation with arguments types only (DB3). Using AST in E11 decreased accuracy compared to all previous rows.

The highest BLEU score for the wrong binary operator was 95.4 (E10), which was similar to the binary operands where the BLEU score reached 95.1 (E10).

\subsection{Heterogeneous Representations (RQ2)}
To address RQ2, we examine whether mixed representations for buggy and fixed code statements assist in learning fixes. We consider E15--E18 here. For swapped arguments, we mix and match representations TF1 and FS2 as they provided the first and second best accuracies of 99.805 and 60.912, respectively. In \autoref{tab:results-swapped-arguments}, E16 achieved one of the highest accuracies of 70.725\% among all. However, E15 yielded the best BLEU score of 76.5, while E16 reached a BLEU score of 75.9. Despite having a substantially lower accuracy, E17 and E18 reached the best positions. 

For the wrong binary operator and operands, representations DB3 and TF1 were employed. In \autoref{tab:results-binary-operator} and \autoref{tab:results-binary-operand}, E16 produced top accuracies of 85.922 and 85.995 along with relatively high BLEU scores of 58.1 and 89.5 for wrong binary operator and operands, respectively. Although E15 has relatively high accuracy, it offers the worst outcome for the position for both wrong binary operator and operands.

\subsection{Embeddings (RQ3)}
To answer RQ3, we compared different embeddings for learning repairs.  For swapped arguments, E10 with abstraction as adopted by Tufano et al. \cite{tufano:tosem:19, Learning-Meaningful-Code-Changes-Via-NMT-2019-7, inproceedings-wild} reached a high accuracy of 99.805. Different embeddings resulted in minor improvements in E19, E20, and E21, as E10 already has high accuracy. GloVe (E20) yielded the best accuracy for swapped arguments. Similarly, GloVe (E20) reached the accuracy of 99.977\% and 99.996\% for the wrong binary operator and operands, respectively, which was the highest accuracy overall. The BLEU score for the embedding experiments (E19--E21) was also relatively high for all bug types.

\renewcommand{\arraystretch}{1.5}
\begin{table}[]
\caption{Results - Usefulness Study for All Bug Types }
\label{tab:results-usefulness}
\resizebox{\textwidth}{!}{%
\begin{tabular}{lccc|cc|cc}
\hline
     & \multicolumn{1}{l|}{}                                                                  & \multicolumn{2}{c|}{\textbf{Swapped Arguments}}                              & \multicolumn{2}{c|}{\textbf{Wrong Binary Operator}}                          & \multicolumn{2}{c}{\textbf{Wrong Binary Operands}}                          \\
\textbf{RID}  & \multicolumn{1}{l|}{\textbf{\begin{tabular}[c]{@{}l@{}}Automatic \\ Patchability\end{tabular}}} & \multicolumn{1}{l}{\textbf{Usefulness}} & \multicolumn{1}{l|}{\textbf{Edit Distance}} & \multicolumn{1}{l}{\textbf{Usefulness}} & \multicolumn{1}{l|}{\textbf{Edit Distance}} & \multicolumn{1}{l}{\textbf{Usefulness}} & \multicolumn{1}{l}{\textbf{Edit Distance}} \\ \hline
WT1  & \cmark & \cellcolor[HTML]{3C78D8}4.4 & 0.0                          & \cellcolor[HTML]{588BDF}3.2 & 0.0                          & \cellcolor[HTML]{3C78D8}4.2 & 0.0                          \\
WT2  & \cmark & \cellcolor[HTML]{3C78D8}4.4 & 27.7                         & \cellcolor[HTML]{588BDF}3.2 & 13.9                         & \cellcolor[HTML]{3C78D8}4.2 & 13.9                         \\
DB1  & \cmark & \cellcolor[HTML]{3C78D8}4.4 & 17.1                         & \cellcolor[HTML]{588BDF}3.2 & 6.7                          & \cellcolor[HTML]{3C78D8}4.2 & 6.7                          \\
DB2  & \cmark & \cellcolor[HTML]{3C78D8}4.4 & 27.1                         & \cellcolor[HTML]{588BDF}3.2 & 22.0                         & \cellcolor[HTML]{3C78D8}4.2 & 22.0                         \\
TF1  & \cmark & \cellcolor[HTML]{3C78D8}4.4 & 30.0                         & \cellcolor[HTML]{588BDF}3.2 & 15.5                         & \cellcolor[HTML]{3C78D8}4.2 & 15.5                         \\
AST1 & \cmark & \cellcolor[HTML]{3C78D8}4.4 & 0.0                          & \cellcolor[HTML]{588BDF}3.2 & 0.0                          & \cellcolor[HTML]{3C78D8}4.2 & 0.0                          \\
AST2 & \cmark & \cellcolor[HTML]{3C78D8}4.4 & 14.4                         & -                           & -                            & -                           & -                            \\ \hline
DB3  & \xmark & \cellcolor[HTML]{6A98E3}3.3 & \cellcolor[HTML]{FF7F1F}34.5 & \cellcolor[HTML]{3C78D8}3.7 & \cellcolor[HTML]{FF6D01}23.4 & \cellcolor[HTML]{598CDF}3.6 & \cellcolor[HTML]{FF6D01}23.4 \\
FS1  & \xmark & \cellcolor[HTML]{83A9E8}2.7 & \cellcolor[HTML]{FF6D01}25.0 & -                           & -                            & -                           & -                            \\
FS2  & \xmark & \cellcolor[HTML]{6695E2}3.4 & \cellcolor[HTML]{FF7611}30.2 & -                           & -                            & -                           & -                            \\
FS3  & \xmark & \cellcolor[HTML]{5E90E0}3.6 & \cellcolor[HTML]{FF730C}28.6 & -                           & -                            & -                           & -                            \\
FS4  & \xmark & \cellcolor[HTML]{6E9BE4}3.2 & \cellcolor[HTML]{FF8022}35.3 & -                           & -                            & -                           & -                            \\
AST3 & \xmark & \cellcolor[HTML]{9CBBEE}2.1 & \cellcolor[HTML]{FF6D01}25.3 & -                           & -                            & -                           & -                            \\
AST4 & \xmark & \cellcolor[HTML]{C9DAF8}1.0 & \cellcolor[HTML]{FCE5CD}88.1 & \cellcolor[HTML]{C9DAF8}1.1 & \cellcolor[HTML]{FCE5CD}70.7 & \cellcolor[HTML]{C9DAF8}1.2 & \cellcolor[HTML]{FCE5CD}70.7

      \\ \hline
\end{tabular}
}
\end{table}

\subsection{Patchability and Usefulness (RQ4)} 
The results for our patchability and usefulness assessment are demonstrated in ~\autoref{tab:results-usefulness}. For each code representation, the table shows its  automatic patchability (binary), usefulness score, and edit distance.  
FS1--FS4 and AST2--AST3 do not apply to wrong binary operator and operands. Edit distance is measuring the effort needed to transform the generated fix suggestion to the actual code.


Token based variants (WT1--WT2, DB1--DB2,  TF1) and AST variants (AST1--AST2) are automatically patchable as they preserve all the relevant information to fix a bug. As lossless transformations could go back to the actual source code, edit-distance does not apply to these representations. Lossy code transformation in token variants (DB3, FS1--FS4) and AST variants (AST3--AST4) are not automatically patchable as they remove information such as function arguments.

Participants found most suggestions to be useful, which is demonstrated by the results from our survey. For swapped arguments in \autoref{tab:results-usefulness},  automatically patchable code representations had the highest perceived usefulness of 4.4. FS3 achieved the second-highest perceived usefulness score of 3.6 even though its edit distance is 28.6. Comparing FS3 and FS1, FS1 has perceived usefulness score of 2.7 while having a lower edit distance of 25.0. Thus closeness in string similarity does not necessarily align with the developer’s perceived usefulness as including additional information could provide helpful debugging aids. For the wrong binary operator, a lossy representation with DeepBugs variant (DB3) yielded the best perceived usefulness of 3.7 compared to the score of 3.2 for automatically patchable code representations. The highest perceived usefulness score for the wrong binary operands reached 4.2 for the automatically patchable representations. 

For mixed representations (E15--E18), the usefulness score stems from the representation of the fixed code. For swapped arguments, wrong binary operator, and wrong binary operands, the best perceived usefulness score achieved by mixed representations are 4.4, 3.7, and 4.2, respectively, which are amongst the best in the corresponding bug types.

\subsection{Overall Best and Worst Representations}
Overall, the best results were achieved in EID(SA) of 20 where the code representation for swapped arguments was TF1 with GloVe embedding and reached the accuracy of 99.964\%, BLEU score of 100, position of 1.0, edit distance of 30.0 and the highest perceived usefulness score of 4.4. TF1 is also automatically patchable. Ignoring AST4, the worst overall results were yielded by EID(Oprt) of 17 where the mix of WT1 and AST1 representations is used, with an accuracy of 10.667\%.

\subsection{Evaluation on Real Bugs}
\label{sec:realworld}

\renewcommand{\arraystretch}{0.8}
\begin{table}[]
\caption{Results - Real Bugs}
\label{tab:real-bugs}
\resizebox{\linewidth}{!}{%
{
\begin{tabular}{cccc}
\toprule
\textbf{EID}         &  \textbf{Accuracy(SA)}      &   \textbf{Accuracy(oprt)}  & \textbf{Accuracy(opnd)}      \\ 
\midrule           
1  & \cellcolor[HTML]{E57D7D}52.17  & \cellcolor[HTML]{56828C}64.86  & \cellcolor[HTML]{BD6F1D}88.57  \\
2  & \cellcolor[HTML]{F0B6B6}30.43  & \cellcolor[HTML]{84A6AD}40.54  & \cellcolor[HTML]{CB8A45}68.57  \\
3  & \cellcolor[HTML]{E57D7D}52.17  & \cellcolor[HTML]{5B8690}62.16  & \cellcolor[HTML]{BB6B18}91.43  \\
4  & \cellcolor[HTML]{E57D7D}52.17  & \cellcolor[HTML]{56828C}64.86  & \cellcolor[HTML]{BD6F1D}88.57  \\
5  & \cellcolor[HTML]{EC9F9F}39.13  & \cellcolor[HTML]{608A93}59.46  & \cellcolor[HTML]{B96712}94.29  \\
6  & \cellcolor[HTML]{EEABAB}34.78  & \multicolumn{1}{c}{-}          & \multicolumn{1}{c}{-}          \\
7  & \cellcolor[HTML]{E99494}43.48  & \multicolumn{1}{c}{-}          & \multicolumn{1}{c}{-}          \\
8  & \cellcolor[HTML]{E99494}43.48  & \multicolumn{1}{c}{-}          & \multicolumn{1}{c}{-}          \\
9  & \cellcolor[HTML]{F4CCCC}21.74  & \multicolumn{1}{c}{-}          & \multicolumn{1}{c}{-}          \\
10 & \cellcolor[HTML]{CC0000}100.00 & \cellcolor[HTML]{608A93}59.46  & \cellcolor[HTML]{D19556}60.00  \\
11 & \cellcolor[HTML]{E57D7D}52.17  & \cellcolor[HTML]{608A93}59.46  & \cellcolor[HTML]{BD6F1D}88.57  \\
12 & \cellcolor[HTML]{E57D7D}52.17  & \multicolumn{1}{c}{-}          & \multicolumn{1}{c}{-}          \\
13 & \cellcolor[HTML]{E99494}43.48  & \multicolumn{1}{c}{-}          & \multicolumn{1}{c}{-}          \\
14 & \cellcolor[HTML]{CC0000}100.00 & \cellcolor[HTML]{134F5C}100.00 & \cellcolor[HTML]{B45F06}100.00 \\
15 & \cellcolor[HTML]{F4CCCC}21.74  & \cellcolor[HTML]{ADC5CA}18.92  & \cellcolor[HTML]{E0B07E}40.00  \\
16 & \cellcolor[HTML]{E06666}60.87  & \cellcolor[HTML]{608A93}59.46  & \cellcolor[HTML]{B96712}94.29  \\
17 & \cellcolor[HTML]{E57D7D}52.17  & \cellcolor[HTML]{D0E0E3}00.00   & \cellcolor[HTML]{FCE5CD}00.00   \\
18 & \cellcolor[HTML]{E57D7D}52.17  & \cellcolor[HTML]{608A93}59.46  & \cellcolor[HTML]{BD6F1D}88.57  \\
19 & \cellcolor[HTML]{CC0000}100.00 & \cellcolor[HTML]{608A93}59.46  & \cellcolor[HTML]{BB6B18}91.43  \\
20 & \cellcolor[HTML]{CC0000}100.00 & \cellcolor[HTML]{608A93}59.46  & \cellcolor[HTML]{B7630C}97.14  \\
21 & \cellcolor[HTML]{CC0000}100.00 & \cellcolor[HTML]{608A93}59.46  & \cellcolor[HTML]{BB6B18}91.43  \\ 
\bottomrule
\end{tabular}
}}
\end{table}

To investigate whether (1) various code representations are useful in fixing real-world bugs, and (2) a model learned using artificial training data can help to repair real bugs, we apply the trained models of all rows in \autoref{tab:matrix-of-experiments} to a set of real-world bugs for all the three types of bugs. These bugs are provided in the DeepBugs paper~\cite{pradel-deepbugs-oopsla-2018}. The number of bugs in each bug pattern is shown in the last column of \autoref{tab:dataset-stat},  
and \autoref{tab:real-bugs} illustrates our results. For swapped arguments, we reach 100\% accuracy using the renaming based abstracted representation (TF1). For the wrong binary operator, 64.86\% is the highest accuracy achieved. 97.14\% is the best accuracy for wrong binary operands as well. Advanced embedding methods help with accuracy, and GloVe is the best among the four embedding methods. We only reported accuracy for the real bugs because the purpose of evaluation on them is to show that a model learned using artificial training data can help to repair real bugs, but complete results of all metrics for real bugs have already been included in the replication package \cite{reptory}.



 \section{Discussion} 
In this section, we will discuss the results presented in Section~\ref{sec:results} and remark on noticeable trends. 

\subsection{BLEU is not always correlated with accuracy.} BLEU score is the first metric measured and provided by the NMT engine itself.
Our experimental results do not always indicate a constant correlation between the BLEU score and accuracy. 
For instance, in \autoref{tab:results-binary-operator} and \autoref{tab:results-binary-operand}, going from E3 to E4 accuracy and BLEU score both increase, however, going from E2 to E3, accuracy drops, but the BLEU score rises. 
The reason might be due to the way the BLEU score is calculated. Suppose that \emph{reference} is the actual fix and \emph{candidate} is what we generate as the fix suggestion; the BLEU score is a fraction where the numerator is the number of words from the candidate that are found in the reference, and the denominator is the total number of words in the candidate. For example, if the reference has four words while the candidate has only two words and both of them are present in the reference the BLEU score will be 100.
Although accuracy is not 100\% here.

\subsection{Bug type affects the functionality of the code representation.}
Various code representations impact accuracy and perceived usefulness differently depending on the bug type. For instance, one situation that shows the role of the bug type can be observed where the effect of abstraction on the wrong binary operator is different from the other two bug types. The reason stems from the nature of this bug, i.e., the operator is important, and removing actual values does not have a negative impact. Moreover, if we only look at the darker rows of the color-coded result tables, we can see that the behaviour of the wrong binary operator and operands is quite similar in terms of accuracy, BLEU score, and position, while it is different in swapped arguments and that is another observation that indicates the essential role of the bug type in code representation for deep-learning. Another example of the role of bug types can be observed in the wrong binary operator results in \autoref{tab:results-usefulness}, where DB3 can reach a relatively high perceived usefulness of 3.7 while it is one of the most abstract representations (\textit{lossy}) we have. We think this originates from the nature of this bug type. To fix binary operator bugs, type information is enough to correct the operator, and the values of operands are not needed. At the same time, removing actual values offers a concise and clean suggestion to developers. These results suggest that depending on the bug type, even a high level of abstraction can be helpful. Although current techniques \cite{hata-learning-to-generate-corrective-patches-2018,inproceedings-coconut,chen-sequencer-2019,mesbah-deepDelta-fse-2019,cure-program-repair-icse-2021,Dinella2020HOPPITY,inproceedings-wild} employ a single code representation for all bug types, our findings suggest that there is no single best code representation applicable to all bug types. This result indicates that neural models with multi-modal representations of source code could be more effective for learning repairs across bug types.


\subsection{More abstraction leads to higher accuracy.}  
Across all the bug patterns, we observe that higher abstractions in code representation help the model in terms of accuracy. For example, for swapped arguments, TF1 in E10 of \autoref{tab:results-swapped-arguments} yielded the best accuracy, where the model did not need to learn the actual values. The same trend holds for the wrong binary operator and operands where E5 produced the best results with the DeepBugs representations with arguments types only (DB3). This is expected as with abstraction, there is limited vocabulary, and learning is easier as the model can exploit the recurring patterns. Also, removing actual values and keeping types, introduces repetitions in the input that helps the model to learn better.

\subsection{More abstraction adversely influences perceived usefulness.}
Developers need to interpret, transform and apply a generated fix suggestion to the buggy code. While abstraction helps accuracy, it can adversely influence the perceived usefulness of fix suggestions. For the swapped arguments bug type, FS1 has the worst perceived usefulness among token based representations in \autoref{tab:results-usefulness}. Similarly, for the wrong binary operands in \autoref{tab:results-usefulness}, DB3 shows the worst perceived usefulness, and we know FS1 and DB3 are among the most abstract representations. For the AST based representations in all bug types, AST with an aggressive level of abstraction (AST4) produces the least useful suggestions.


\subsection{Lossy representations can still be helpful.}
In some code representations, such as word tokenization (WT1), a generated fix suggestion possesses all the information needed to fix the bug. At the same time, some fixes generated with lossy representations, such as function signature (FS1), cannot be applied to the buggy code directly because they have lost some information, and it is impossible to generate the exact valid code from the fix suggestion. We refer to the lossless code representations that are convertible to the actual code as automatically patchable code representations. Similarly, lossy code representations are not automatically patchable.
Therefore, we need another step to determine if the lossy code representations are completely useless or not. As a result, the developer study is performed.
Our user study finds that lossy fix suggestions might still be valuable during debugging tasks for developers.
Consider the following real-world example from our experiment:

\begin{lstlisting} [caption={Example of a fix suggestion for \autoref{lst:swappedArgumentExample} in FS3 representation.},captionpos=b,label={lst:fixexample}]
\\Actual buggy version:
    setTimeout ( delay , fn )
\\Actual fixed version:
    setTimeout ( fn , delay ) 
\\The developer sees the following as a fix suggestion:
    setTimeout ( ID function , ID number)
\end{lstlisting}

This inference was helpful to developers because it provides a useful debugging cue about how to invoke a function with the correct argument order.
Lossy code representations abstract away some information, leading to higher accuracy while they are not automatically patchable. Therefore, the high accuracy might be misleading, and that is where the perceived usefulness comes into play.
For instance, we can assume the focus is not on the accuracy, and we use the perceived usefulness score to select the best code representation. Among the lossy code representations, function signature with LIT/ID (FS3) produces the highest perceived usefulness score for the swapped arguments bug type in \autoref{tab:results-usefulness}. Also, among the lossy code representations, DeepBugs with arguments types only (DB3) shows the highest perceived usefulness for both wrong binary operator and operands. The lossless code representations share the same perceived usefulness per bug type because all of them could be automatically converted into the actual valid code. Generally, the perceived usefulness of lossless code representations is higher than lossy ones except for the wrong binary operator bug type. The reason stems from the nature of this bug where the incorrect operator is the bug, and it does not matter if the operands are abstracted.
Our work highlights that in addition to accuracy, the usefulness of the model output has to be assessed, especially when a patch is not automatically convertible. Indeed, the purpose of doing the user study was exactly to address this issue and observe if the abstract fix suggestions (\textit{lossy}) help developers.

\subsection{String relatedness is not aligned with perceived usefulness.} 
As our results in \autoref{tab:results-usefulness} show, for FS1 and FS3 code representations, the edit distance is 25.0 and 28.6, respectively. On the other hand, their perceived usefulness scores are 2.7 and 3.6, respectively. Even though FS1 was closer to the actual fix, it yielded a lower perceived usefulness score because additional information in FS3 helps developers with extra debugging aids. A similar observation could be made for the DB3 and FS1 code representations. DB3 has a higher usefulness score of 3.3 even though FS1 was closer to the actual fix in terms of the edit distance. Thus, the perceived usefulness of a code representation from a real developer's perspective does not align with the effort needed to convert the generated fix suggestion to the actual code. One code representation might be more useful while it takes more effort to be transformed into the actual code.

\subsection{Heterogeneous representation could outperform homogeneous code representation.} 
Heterogeneous code representation could be effective across all three bug types. Mixed representations, in which fixed code is more abstracted than buggy code, could yield higher accuracy than homogeneous representations. For swapped arguments, E16 presents a heterogeneous code representation comprised of FS2 as buggy code and TF1 as the fixed code representation. FS2 representation is enriched with function signature and position anchors, whereas TF1 abstracts away method names, variables, and literals and does not have additional information. Hence, TF1 has less information and is more abstract compared to FS2. When E16 (FS2 $\rightarrow$ TF1) is compared to E7 (FS2 $\rightarrow$ FS2), E16 has a greater accuracy of 70.725, while E7 has a lower accuracy of 60.912. 
As we go from a less abstract code representation (more information) to a more abstract code representation (less information), the likelihood of improving inference accuracy increases. However, if we go from a more abstract buggy code representation, we do not keep all the fix ingredients for less abstract representation during inference and accuracy drops. E16 has higher accuracy than E7 because TF1 is more abstract than FS2, and we observe that a heterogeneous code representation outperforms a homogeneous code representation.
The same trend remains valid for the wrong binary operator and operands. Since our user study showed that developers could tolerate some level of abstraction, heterogeneous representation could be chosen over homogeneous representation (with a more abstract code representation for the fixed version of code) to improve accuracy.

\subsection{Ordering matters in heterogeneous code representations.} 
In mixed representation, using a lower abstraction (more information) for the buggy code along with a higher abstraction for the fixed code (less information) improves the accuracy. For swapped arguments, E15 goes from TF1 (more abstract, less information) to FS2 (less abstract, more information), yields an accuracy of 46.666. However, E16 employs the inverse order and goes from FS2 to TF1, resulting in a significantly higher accuracy of 70.725.
Similarly, E18 (AST1 $\rightarrow$ WT1) yields an accuracy of 20.690, which is higher than 19.913 in E17 (WT1 $\rightarrow$ AST1). The same pattern holds for wrong binary operator and operands, indicating that ordering of code representations play a role in heterogeneous code representations.

\subsection{Advanced embedding methods improve learning.}
Word2Vec-SG does not seem to affect the accuracy in comparison with Word2Vec-CBOW. GloVe yields the best accuracy for all the bug types in our study. This could be explained by the fact that GloVe focuses on word co-occurrences over the whole corpus, and its embeddings relate to the probabilities that two words appear together. Therefore, it produces more accurate embeddings for source code. FastText does not improve the accuracy as much as GloVe does.

\subsection{Correct patches can be generated for most bugs in both test and real datasets.}
Although we conducted a controlled experiment and compared various code representations in this study, the overall results indicate that deep-learning can learn fixes for name-based bugs. With the right code representation and embedding, for swapped arguments, wrong binary operator, and wrong binary operands, correct patches can be generated for more than 97\% of the bugs in both test and real datasets with one exception where in the real bugs dataset, wrong binary operator bugs are fixed with the accuracy of 64.86\%.

\subsection{Controlled experiments in DL systems are challenging.}
The performance and accuracy metrics for a learning-based model are influenced by the hyperparameters. While numerous hyperparameters can be tuned for different code representations individually, this is not feasible in a controlled experiment where the intent is to keep everything constant and observe the effect of changing code representations. In this work, we encountered a number of challenges pertaining to conducting controlled experiments for deep learning-based analysis, namely (1) there is a vast number of factors that need to be controlled for, such as neural network architectures, values for hyper-parameters, and embedding methods (e.g., Word2Vec, GloVe, FastText), in addition to various code abstractions and representations; the permutations of these variables populate a large optimization learning matrix, unlike anything we have seen in the literature thus far; (2) training models on large datasets is time-consuming and computationally expensive. Overall, conducting controlled experiments to compare deep-learning models is a challenging task due to the vast number of variables involved. However, we believe it is an essential vehicle for proper empirical comparison in such an emerging and fast-changing area where new ideas are proposed increasingly.

\subsection{Threats to Validity}

This section outlines how we mitigated potential limitations that may have biased our findings.

\subsubsection{External validity} 

\header{Choice of programming languages}
In this work, we study the impact of code representation in JavaScript. Since we do not rely on language-specific features, our findings should apply to other programming languages, although more controlled experiments are needed in other languages to empirically assess this.

\header{Selection of bug types}
Another concern could be that this study is limited to specific bug patterns. We attempted to mitigate the generalizability threat by incorporating three bug types used widely in the literature~\cite{pradel-deepbugs-oopsla-2018, inproceedings-5, Karampatsis2020HowOD, wainakh-semantic-representations-of-source-code-2019, Briem2020OffSideLT}, which are also among the most prevalent JavaScript bug categories~\cite{hanam-discovering-bug-patterns-fse-2016,articlepan}. In addition, many existing learning-based bug detection tools are trained and tested on these bug-types~\cite{pradel-deepbugs-oopsla-2018, Briem2020OffSideLT, sellik:off-by-one-msr:21}. As a result, studying these bug categories is both representative and significant. However, we do not claim or generalize that empirical results would be directly applied to more complex bug types, especially in conjunction with more different deep learning models such as Transformers~\cite{transformer:neurips:2017} and Graph Neural Network~\cite{gnn-survey:2021}. However, we believe our work is the first step toward  controlled experiementation and our framework, \toolname, is designed to allow the addition of other bug patterns for experimentation as described in details in Section \ref{ssec:datacollection}. 

\header{Model selection} We opted for the NMT model, which is well studied in the recent literature \cite{tufano:tosem:19, inproceedings-wild, Learning-Meaningful-Code-Changes-Via-NMT-2019-7, chen-sequencer-2019, hata-learning-to-generate-corrective-patches-2018}. We did not compare our model(s) with other ML models such as Transformers~\cite{transformer:neurips:2017}, Graph Neural Network~\cite{gnn-survey:2021}, or Tree-LSTM~\cite{tree-lstm:iclr:2020}. The goal of this work is not to compare different models with the best performance; rather, we evaluate the effect of code representations in a controlled setting and for that, we have adopted a widely used NMT-based architecture. Moreover, adding more models is computationally costly (for training the models on different code representations) but may lead to other interesting observations. For future comparisons, our replication package can be used as a starting point.

\header{Contextual information, graphical representations of code and learning models} Context pertaining to the buggy statement could play a role in the representation, which we have not considered in this study. Context can be extracted in various levels of granularity such as the buggy statement~\cite{pradel-deepbugs-oopsla-2018, hata-learning-to-generate-corrective-patches-2018}, surrounding statements, enclosing function~\cite{inproceedings-coconut,watson-learning-meaningful-assert-statements-icse-2020}, class~\cite{inproceedings-wild,chen-sequencer-2019}, enclosing file~\cite{sakib-summarization-msr-2020}, or encapsulating AST subtrees~\cite{subtree,inproceedings-6}.

After choosing the granularity of context from this wide array of options, there are numerous ways of modelling source code to exploit syntactic and semantic relations. For example, an abstract syntax tree could represent the syntactical structure, whereas semantics information could be derived from control flow and data flow graphs. Furthermore, there could be numerous variations of combining the syntactical and semantics representations. One can fuse semantic information, e.g., control flow with the abstract syntax tree, data flow with the abstract syntax tree, or both. 

Intuitively, graph neural networks could effectively model such representations and learn from graphical representations of code. There are different graph neural networks~\cite{zhou:gnnsurvey} with varying levels of expressiveness. In the context of GNN, expressiveness means their ability to represent different graph structures~\cite{sato:expressiveness-gnn:20}. GraphSage~\cite{hamilton:graphsage:17
}, for example, can scale well to large source code input but is not expressive enough since, when aggregating information, it does not use all of the node neighbours but only samples some of them. On the other hand, Graph Attention Network (GAT)~\cite{velickovic:gat:18} uses the information from all its neighbours along with the attention mechanism. There are also variants that differ in how node and edge features are leveraged.

Understanding the impact of granularity of contextual information, context representation using different graphical representations, and the interplay with varying deep-learning models is a non-trivial task. We believe future work could explore the possibility of combining syntactic information with deeper contextual information, which could then be used to more comprehensively test the capabilities of the different learning models to further understand the intricacies of doing learning based program repair. We believe our work is the first to indicate the intricacies and challenges of performing controlled experiments on learning-based task on source code and the importance of such endeavours. Our replication package is available and could be expanded in future studies on context.

\header{Contextual embedding} 
The use of embeddings to represent textual words as vectors has emerged as a powerful technique for associating meanings with words and comparing words for similarity based on such meanings. In this work, we focus on static word representation models to obtain embeddings that are not dependent on their surrounding code tokens or code statements (context). Recently, contextual embedding models~\cite{mccann:context-vector:17,peters:elmo:18,devlin:bert:19} in natural language processing have been adopted, which can embed contextual information as well as word representations, making them more expressive than static word representations. Subsequently, the notion of contextual embedding is applied to source code and contextualized embeddings such as  CodeBERT~\cite{feng-etal-2020-codebert},  CuBERT~\cite{inproceedings-5}, and SCELMo~\cite{karampatsis:scelmo:20} have shown promise in many different source code tasks. In contrast in this work, we compare four static word embeddings namely, Word2Vec-CBOW, Word2Vec-SG, FastText, and GloVe, which is consistent with prior work~\cite{wainakh-semantic-representations-of-source-code-2019}. A gradual progression would be to devise a carefully designed comparative study on contextual embedding in conjunction with different granularities of contextual information. 

\header{Mutated dataset}
Our work performs mutations to generate faulty examples from likely correct code that may not represent real-world bugs. However, the notion of artificially creating likely incorrect code relates to mutation testing \cite{jia-survey-mutation-testing}, and it is shown that mutants are quite representative of real bugs in general \cite{Just2014AreMA, Andrews2005IsMA, howclose}. Furthermore, our evaluation shows that models trained with artificially seeded bugs are also effective for repairing real-world bugs.

\header{Real bugs dataset}
The dataset of real bugs is small. The goal here is to evaluate whether trained models could be effective on real-world bugs. The dataset for real bugs is derived from DeepBugs \cite{pradel-deepbugs-oopsla-2018} which performs bug prediction. Here, we present a generative model for a program repair task to produce fix suggestions on a real bug dataset. Although the dataset of real bugs is small, we obtain similar results. 

\header{Code representations and bug types}
We investigate function signature-related representations, given that they do not apply to all selected bug patterns. In terms of applicability, two categories of code representations exist. Some representations are applicable to all bug types \cite{inproceedings-coconut} such as word tokenization, but another category of code representations has been used in the literature, which is specific to a particular bug type \cite{inproceedings-learned-heuristics}. To make the experiment more comprehensive, we involved both kinds of representations.

\subsubsection{Internal validity} 

\header{Perceived usefulness}
Another potential threat, which is currently present in many learning-based techniques, would be focusing only on the \textit{accuracy} metric. While the more abstracted representation (\textit{lossy}) reaches a higher accuracy, it can produce a less useful bug fix. To minimize this threat, we conducted a user study to assess the perceived usefulness of the suggestions, in addition to accuracy.

\header{Selection criterion for mixed representations}
We decided to use accuracy for the selection since it is an objective, repeatable indicator widely used in the literature \cite{chen-sequencer-2019, Dinella2020HOPPITY, gupta-deepfix-aaai-2017, Hajipour2019SampleFixLT, inproceedings-coconut, marginean-sapFix-icse-2019, mesbah-deepDelta-fse-2019, Learning-Meaningful-Code-Changes-Via-NMT-2019-7}. We discuss that edit distance is not aligned with the developer perceived usefulness. The BLEU score is not a good proxy either, as observed in other recent studies \cite{ding-patching-as-translation-ase-2020}. Also, the usefulness metric could be perceived as subjective due to the nature of the user study design. If we combine all dependent variables using the same weight i.e., ${\scriptstyle Accuracy * BLEU Score * (1/Position) * Usefulness * (1/Edit Distance)} $, for the swapped arguments, WT1 would yield the best result. AST1 would become the second best representation because of the lower edit distance. As we combine all the scores, the first best and second-best pair would change for the mixed representations (EID15 and EID16) from (TF1, FS2) to (WT1, AST1) in \autoref{tab:results-swapped-arguments}. However, if we do not consider edit distance in the combined score for automatically patchable code representations, TF1 would still yield the best result. Regardless, this would not impact our findings and discussions on mixed representation, as to why mixed representation could be effective and why ordering is important for mixed representations.


\header{Hyperparameter tuning}
One approach for hyperparameter optimization could be to tune each code representation from \autoref{tab:matrix-of-experiments}. Finding an optimal setting can be  computationally expensive due to the large search space of all available hyper-parameters, different variations of code representations and different bug types. Overall, this would incur an exorbitantly high cost for training numerous models. Furthermore, employing different hyperparameters would introduce additional variability into our controlled experiment. Hence, we opted for tuning for the code representation WT1, which is closest to the actual code. We verified that the tuned parameters with the least abstracted code representation work better than the random sets of hyperparameter values for other code representations. As a result, we kept the hyperparameter values constant throughout the entire experimental matrix in our controlled experiment.

\header{Synthesized types}
For function signature related representations, we have incorporated synthesized types for variables. These synthesized types are not necessarily always the actual type, which might raise a concern. However, the goal here was not to correctly infer types in JavaScript; our goal was to observe what would happen if we abstracted out and used higher-level information from the code. The synthesized types for arguments are added consistently such that argument types are always the same for a particular function.

\header{Reproducibility}
In terms of the reproducibility of our results, we have made \toolname~\cite{reptory}, our framework and dataset, available, along with instructions to use it for replication or future experimental comparisons of other code representations.

\section{Related Work}

We will discuss related work in this section. Firstly, the name-related program analysis summarises previous research in the topic. We present different embeddings applied to source code processing tasks. Then we discuss the translation and edit-based models for learning-based program repair.  Followed to that, we outline existing comparative research on ML for SE. Since we are conducting a controlled experiment and comparing various code representations, we are interested to review previous research that has compared different components involved in the learning-based source code analysis.

\header{Name-related program analysis} There is a wide array of research on name-related program analysis~\cite{pradel-deepbugs-oopsla-2018,argument-selection-defects-oopsla-2017,detecting-swapped-arguments-scam-2020,dl-identify-suspicious-return-saner-2020,argument-defects-icse-2016,identifier-renamings-msr-2011,argument-defects-issta-2011,arguments-defects-tse-2013,allamanis-name-based-analysis-method-naming-fse-2015,pandita-name-based-method-specification-from-naming-2012,pandita-name-based-inferring-constraints-from-naming-2016}. There have been attempts to use the rich information embedded in source code identifiers to perform method naming~\cite{allamanis-name-based-analysis-method-naming-fse-2015}, infer API specifications~\cite{pandita-name-based-method-specification-from-naming-2012,pandita-name-based-inferring-constraints-from-naming-2016}, and detect argument defects~\cite{argument-selection-defects-oopsla-2017,argument-defects-icse-2016}. Recent studies also applied the learning-based approach to detect name-based bugs~\cite{pradel-deepbugs-oopsla-2018,dl-identify-suspicious-return-saner-2020}. To the best of our knowledge, this is the first work on a generative task on name-related bugs, rather than focusing on a predictive model. 

\header{Embeddings on source code processing tasks} Word embeddings provide a way to represent words as vectors such that words that are closer to the vector spaces are related in meaning. In natural language processing word embedding techniques different embedding techniques, Word2Vec, Glove, and FastText, have been proposed. Word embedding techniques like Word2Vec have been heavily adopted in learning-based source code processing. Learning Word2vec embedding from token sequences is applied to different source code processing tasks, namely for bug detection~\cite{pradel-deepbugs-oopsla-2018}, type prediction~\cite{nl2type}, or vulnerability detection~\cite{vulnerability:tang:2020}. Chen et al.~\cite{chen:embeddings-lit:19} provide a more in-depth investigation of source code embeddings. In natural language processing, contextual embedding~\cite{liu:contextual-survey:20} like BERT~\cite{devlin:bert:19}, and ELMo~\cite{peters:elmo:18} is proposed where instead of static word embedding, higher-quality embeddings could be learned from the usage context. Recently, these contextualized word embeddings are also adopted in source code processing~\cite{inproceedings-5, feng-etal-2020-codebert,karampatsis:scelmo:20}. In this work, we focus on static word embedding and compare 4 embedding techniques. Contextualized word embedding with varying granularities of contextual information is a logical progression of this work that could be built on the \toolname framework.

\header{Deep learning-based program repair} Recently, neural machine translation (NMT) based architectures have been employed to generate bug fixes~\cite{hata-learning-to-generate-corrective-patches-2018,tufano:tosem:19, Learning-Meaningful-Code-Changes-Via-NMT-2019-7, inproceedings-wild}. SequenceR~\cite{chen-sequencer-2019} employed the sequence-to-sequence model in conjunction with the copy mechanism to further improve the effectiveness of NMT based model. To improve the repair technique, CoCoNuT~\cite{inproceedings-coconut} adopts CNN to encode the faulty method and generates the patch token by token. Techniques from this stream of work consider the bug-fix learning problem as sequence-based code generation and harness the power of sequence-to-sequence based models. In recent work, Ding et al.~\cite{ding-patching-as-translation-ase-2020} looked critically at using a neural translation-based approach for the program repair task. In contrast to these translation-based models, several edit-based techniques have been proposed which perform code edits on the buggy code. Instead of sequence-based correct code generation, edit-based approaches~\cite{zhao:neural-edits:icrl:19, chakraborty-codit-2018} generate the sequence of edits given the buggy piece of code. Instead of token-based representation of source code, Hoppity~\cite{Dinella2020HOPPITY} performs edit in the abstract syntax tree (AST). Hoppity trains a graph edit model using GNN-based architecture and iteratively transforms the AST of the buggy code into the AST of the fixed code. 

The goal of these approaches is to propose different program repair techniques using a varied level of code representation, embedding, and learning models. Instead in this work, we focus on a controlled setting wherein we examine the effect of code representation on the learning model in a systematic manner.

\header{Comparative studies on ML for SE} A few studies have attempted to understand the effect of changing factors, such as code embedding or neural network architecture, on various software engineering tasks such as bug detection, method name prediction, and code cloning. Wainakh et al. \cite{wainakh-semantic-representations-of-source-code-2019} compare five different identifier embeddings in the source code and show to what extent each of them captures the relatedness, similarity, and contextual similarity between identifiers. COSET \cite{wang-coset-2019} offers a benchmarking framework consisting of a diverse dataset of programs in the source-code format, labeled by human experts. It trains four different neural network architectures proposed for software classification tasks and uses prediction accuracy to measure the classification results. Kanade et al. \cite{inproceedings-5} use pre-trained contextual embeddings of source-code as well as Word2Vec for five classification tasks and compare the performance of these two approaches.

To predict if a piece of code is clean or vulnerable, Li et al. \cite{article-4} compare six different neural network architectures and five various embedding methods. Most of the previous studies focus on predictive tasks and classification problems. A previous work~\cite{Learning-Meaningful-Code-Changes-Via-NMT-2019-7} that concentrates on generative tasks, investigates whether NMT can be used to learn code transformations on a method-level granularity. As part of the evaluation, this study compared the effect of using small and medium-sized methods on the accuracy of the generated code.

Other studies compare different APR tools as part of their evaluation section~\cite{inproceedings-coconut, Liu2019YouCF, li-dlfix-2020,cure-program-repair-icse-2021}. Their target is not to do a controlled experiment but to propose a method and compare it with existing approaches.  Unlike our study, none of these studies conducts experiments in a controlled way and explores the effect of different changing factors on learning performance as well as on each other. Instead, they take some factors, mutate them independently and observe the results.

A study conducted by Tufano et al. \cite{inproceedings-6} applied four different representations of source-code to detect code similarities. This work is the closest to ours, where the impact of code representation is studied, albeit for a completely different task. In this paper, we consider three changing factors, code representations, mixed representations, and embeddings. The main focus is on code representations, and once the code representation is fixed, we study how mixed representations and embeddings affect code representations to generate fix suggestions. Therefore, the novelties of this work are the combination of three ideas: (1) coverage of the main code representation categories, namely token based and AST-based, (2) comparison in a controlled way where three independent variables are involved, and (3) comparison for a generative task, namely, patch generation.

\section{Conclusion}
Learning-based software engineering is gaining traction in research and development. However, there is a lack of clear understanding of how to best represent source code for learning-based tasks. In this paper, we studied the effects of code representations and embedding methods in 21 different ways on name-based program repair tasks through a controlled experiment. We demonstrated that string relatedness is not a good proxy to assess the patchability of inferred fixes. We qualitatively evaluated the perceived usefulness of a code representation in fix suggestions, which is overlooked in the learning-based repair literature. Our results indicate that a higher abstraction level increases accuracy while decreasing perceived usefulness depending on the bug type. However, lossy representation could still yield useful debugging hints to the developers. The notion of heterogeneous code representation, proposed in this paper for a sequence to sequence repair task, can outperform homogeneous code representations. The results of our user study emphasize the need to develop learning-based repair tools with practitioners in mind. 
The best accuracy achieved for swapped arguments, wrong binary operator and wrong binary operands are 99.964\% (\autoref{tab:results-swapped-arguments}, EID(SA)-20), 99.977\% (\autoref{tab:results-binary-operator}, EID(Oprt)-20) and 99.996\% (\autoref{tab:results-binary-operand}, EID(Opnd)-20), respectively, where each test dataset has more than 120k data points in \autoref{tab:dataset-stat}. 
\toolname can be used as a benchmarking framework to perform comparative studies when considering or proposing new code representations for learning-based tasks.

We intend to expand our work to cover other bug types and develop a configurable infrastructure in the future. Such an infrastructure can be fed by various independent variables, then run experiments and generate results based on desired dependent variables. Developing an automatic evaluation metric for source code is another promising idea to follow. Another possibility is to combine context with single-statement bugs and contextual embedding techniques to investigate their impact on repair tasks.


%
%

\bibliographystyle{spmpsci}      
\bibliography{automated-javascript-repair}   

\begin{thebibliography}{100}
\providecommand{\url}[1]{{#1}}
\providecommand{\urlprefix}{URL }
\expandafter\ifx\csname urlstyle\endcsname\relax
  \providecommand{\doi}[1]{DOI~\discretionary{}{}{}#1}\else
  \providecommand{\doi}{DOI~\discretionary{}{}{}\begingroup
  \urlstyle{rm}\Url}\fi

\bibitem{Ahmed2019LearningLP}
Ahmed, T., Devanbu, P., Hellendoorn, V.J.: Learning lenient parsing {\&} typing
  via indirect supervision.
\newblock Empirical Software Engineering \textbf{26}(2) (2021)

\bibitem{allamanis2019adverse}
Allamanis, M.: The adverse effects of code duplication in machine learning
  models of code.
\newblock In: ACM SIGPLAN International Symposium on New Ideas, New Paradigms,
  and Reflections on Programming and Software, Onward! 2019, p. 143–153.
  Association for Computing Machinery (2019)

\bibitem{allamanis-name-based-analysis-method-naming-fse-2015}
Allamanis, M., Barr, E.T., Bird, C., Sutton, C.: Suggesting accurate method and
  class names.
\newblock In: 10th Joint Meeting on Foundations of Software Engineering,
  ESEC/FSE 2015, p. 38–49. Association for Computing Machinery (2015)

\bibitem{Andrews2005IsMA}
Andrews, J.H., Briand, L.C., Labiche, Y.: Is mutation an appropriate tool for
  testing experiments?
\newblock In: 27th International Conference on Software Engineering, p.
  402–411. Association for Computing Machinery (2005)

\bibitem{Scott2019GetafixLT}
Bader, J., Scott, A., Pradel, M., Chandra, S.: Getafix: Learning to fix bugs
  automatically.
\newblock Proc. ACM Program. Lang. \textbf{3}(OOPSLA) (2019)

\bibitem{bahdanau-attention-iclr-2015}
Bahdanau, D., Cho, K., Bengio, Y.: Neural machine translation by jointly
  learning to align and translate.
\newblock In: 3rd International Conference on Learning Representations, {ICLR}
  2015 (2015)

\bibitem{recommendations-for-tuning-bengio-2012}
Bengio, Y.: Practical recommendations for gradient-based training of deep
  architectures.
\newblock In: Neural Networks: Tricks of the Trade - Second Edition,
  \emph{Lecture Notes in Computer Science}, vol. 7700, pp. 437--478. Springer
  (2012)

\bibitem{Bielik2016PHOGPM}
Bielik, P., Raychev, V., Vechev, M.: Phog: Probabilistic model for code.
\newblock In: 33rd International Conference on International Conference on
  Machine Learning - Volume 48, ICML'16, p. 2933–2942. JMLR.org (2016)

\bibitem{article-fasttext1}
Bojanowski, P., Grave, E., Joulin, A., Mikolov, T.: Enriching word vectors with
  subword information.
\newblock Transactions of the Association for Computational Linguistics
  \textbf{5}, 135--146 (2017)

\bibitem{Briem2020OffSideLT}
Briem, J.A., Smit, J., Sellik, H., Rapoport, P., Gousios, G., Aniche, M.:
  Offside: Learning to identify mistakes in boundary conditions.
\newblock In: IEEE/ACM 42nd International Conference on Software Engineering
  Workshops, ICSEW'20, p. 203–208. Association for Computing Machinery (2020)

\bibitem{Brody2020ASM}
Brody, S., Alon, U., Yahav, E.: A structural model for contextual code changes.
\newblock Proc. ACM Program. Lang. \textbf{4}(OOPSLA) (2020)

\bibitem{chakraborty-codit-2018}
Chakraborty, S., Ding, Y., Allamanis, M., Ray, B.: Codit: Code editing with
  tree-based neural models.
\newblock IEEE Transactions on Software Engineering pp. 1--1 (2020)

\bibitem{chandra:angelic-debugging:icse:11}
Chandra, S., Torlak, E., Barman, S., Bodik, R.: Angelic debugging.
\newblock In: 33rd International Conference on Software Engineering, ICSE '11,
  p. 121–130. Association for Computing Machinery (2011)

\bibitem{chen-sequencer-2019}
{Chen}, Z., {Kommrusch}, S.J., {Tufano}, M., {Pouchet}, L., {Poshyvanyk}, D.,
  {Monperrus}, M.: Sequencer: Sequence-to-sequence learning for end-to-end
  program repair.
\newblock IEEE Transactions on Software Engineering pp. 1--1 (2019)

\bibitem{unknown-chen}
Chen, Z., Monperrus, M.: The remarkable role of similarity in redundancy-based
  program repair.
\newblock arXiv preprint arXiv:1811.05703  (2018)

\bibitem{chen:embeddings-lit:19}
Chen, Z., Monperrus, M.: A literature study of embeddings on source code.
\newblock arXiv preprint arXiv:1904.03061  (2019)

\bibitem{devlin:bert:19}
Devlin, J., Chang, M., Lee, K., Toutanova, K.: {BERT:} pre-training of deep
  bidirectional transformers for language understanding.
\newblock In: North American Chapter of the Association for Computational
  Linguistics: Human Language Technologies, {NAACL-HLT}, pp. 4171--4186.
  Association for Computational Linguistics (2019)

\bibitem{article-semantic-code-repair}
Devlin, J., Uesato, J., Singh, R., Kohli, P.: Semantic code repair using
  neuro-symbolic transformation networks.
\newblock arXiv preprint arXiv:1710.11054  (2017)

\bibitem{Dinella2020HOPPITY}
Dinella, E., Dai, H., Li, Z., Naik, M., Song, L., Wang, K.: Hoppity: Learning
  graph transformations to detect and fix bugs in programs.
\newblock In: 8th International Conference on Learning Representations, {ICLR}
  2020. OpenReview.net (2020)

\bibitem{ding-patching-as-translation-ase-2020}
Ding, Y., Ray, B., Devanbu, P., Hellendoorn, V.J.: Patching as translation: The
  data and the metaphor.
\newblock In: 35th IEEE/ACM International Conference on Automated Software
  Engineering, ASE '20, p. 275–286. Association for Computing Machinery
  (2020)

\bibitem{dolan-lava-2016}
{Dolan-Gavitt}, B., {Hulin}, P., {Kirda}, E., {Leek}, T., {Mambretti}, A.,
  {Robertson}, W., {Ulrich}, F., {Whelan}, R.: Lava: Large-scale automated
  vulnerability addition.
\newblock In: 2016 IEEE Symposium on Security and Privacy (SP), pp. 110--121
  (2016)

\bibitem{inproceedings-dynamoth}
{Durieux}, T., {Monperrus}, M.: Dynamoth: Dynamic code synthesis for automatic
  program repair.
\newblock In: 2016 IEEE/ACM 11th International Workshop in Automation of
  Software Test (AST), pp. 85--91 (2016)

\bibitem{identifier-renamings-msr-2011}
Eshkevari, L.M., Arnaoudova, V., Di~Penta, M., Oliveto, R., Gu\'{e}h\'{e}neuc,
  Y.G., Antoniol, G.: An exploratory study of identifier renamings.
\newblock In: 8th Working Conference on Mining Software Repositories, MSR '11,
  p. 33–42. Association for Computing Machinery (2011)

\bibitem{feng-etal-2020-codebert}
Feng, Z., Guo, D., Tang, D., Duan, N., Feng, X., Gong, M., Shou, L., Qin, B.,
  Liu, T., Jiang, D., Zhou, M.: {C}ode{BERT}: A pre-trained model for
  programming and natural languages.
\newblock In: Findings of the Association for Computational Linguistics: EMNLP
  2020, pp. 1536--1547. Association for Computational Linguistics (2020)

\bibitem{reptory}
The impact of code representation on learning-based repair replication package.
\newblock \url{https://github.com/annon-reptory/reptory} (2021)

\bibitem{howclose}
Gopinath, R., Jensen, C., Groce, A.: Mutations: How close are they to real
  faults?
\newblock In: 2014 IEEE 25th International Symposium on Software Reliability
  Engineering, ISSRE '14, p. 189–200. IEEE Computer Society (2014)

\bibitem{article-deep-reinforcement}
Gupta, R., Kanade, A., Shevade, S.K.: Deep reinforcement learning for
  programming language correction.
\newblock arXiv preprint arXiv:1801.10467  (2018)

\bibitem{Gupta2017DeepFixFC}
Gupta, R., Pal, S., Kanade, A., Shevade, S.: Deepfix: Fixing common c language
  errors by deep learning.
\newblock In: Thirty-First AAAI Conference on Artificial Intelligence, AAAI'17,
  p. 1345–1351. AAAI Press (2017)

\bibitem{gupta-deepfix-aaai-2017}
Gupta, R., Pal, S., Kanade, A., Shevade, S.: Deepfix: Fixing common c language
  errors by deep learning.
\newblock In: Thirty-First AAAI Conference on Artificial Intelligence, AAAI'17,
  p. 1345–1351. AAAI Press (2017)

\bibitem{Hajipour2019SampleFixLT}
Hajipour, H., Bhattacharyya, A., Fritz, M.: Samplefix: Learning to correct
  programs by efficient sampling of diverse fixes.
\newblock In: NeurIPS 2020 Workshop on Computer-Assisted Programming (2020)

\bibitem{hamilton:graphsage:17}
Hamilton, W.L., Ying, R., Leskovec, J.: Inductive representation learning on
  large graphs.
\newblock In: 31st International Conference on Neural Information Processing
  Systems, NIPS'17, p. 1025–1035. Curran Associates Inc. (2017)

\bibitem{hanam-discovering-bug-patterns-fse-2016}
Hanam, Q., Brito, F.S.d.M., Mesbah, A.: Discovering bug patterns in javascript.
\newblock In: 24th ACM SIGSOFT International Symposium on Foundations of
  Software Engineering, FSE 2016, p. 144–156. Association for Computing
  Machinery (2016)

\bibitem{sakib-summarization-msr-2020}
Haque, S., LeClair, A., Wu, L., McMillan, C.: Improved automatic summarization
  of subroutines via attention to file context.
\newblock In: 17th International Conference on Mining Software Repositories,
  MSR '20, p. 300–310. Association for Computing Machinery (2020)

\bibitem{hartmann:suggesting-solutions:chi:2010}
Hartmann, B., MacDougall, D., Brandt, J., Klemmer, S.R.: What would other
  programmers do: Suggesting solutions to error messages.
\newblock In: SIGCHI Conference on Human Factors in Computing Systems, CHI '10,
  p. 1019–1028. Association for Computing Machinery (2010)

\bibitem{hata-learning-to-generate-corrective-patches-2018}
Hata, H., Shihab, E., Neubig, G.: Learning to generate corrective patches using
  neural machine translation.
\newblock arXiv preprint arXiv:1812.07170  (2018)

\bibitem{deepcom}
Hu, X., Li, G., Xia, X., Lo, D., Jin, Z.: Deep code comment generation.
\newblock In: 26th Conference on Program Comprehension, ICPC '18, p. 200–210.
  Association for Computing Machinery (2018)

\bibitem{jeffrey:bugfix:icpc:2009}
Jeffrey, D., Feng, M., Gupta, N., Gupta, R.: Bugfix: A learning-based tool to
  assist developers in fixing bugs.
\newblock In: 2009 IEEE 17th International Conference on Program Comprehension
  (ICPC 2009). IEEE Computer Society (2009)

\bibitem{jia-survey-mutation-testing}
{Jia}, Y., {Harman}, M.: An analysis and survey of the development of mutation
  testing.
\newblock IEEE Transactions on Software Engineering \textbf{37}(5), 649--678
  (2011)

\bibitem{cure-program-repair-icse-2021}
Jiang, N., Lutellier, T., Tan, L.: Cure: Code-aware neural machine translation
  for automatic program repair.
\newblock In: 2021 IEEE/ACM 43rd International Conference on Software
  Engineering (ICSE), pp. 1161--1173. IEEE Computer Society (2021)

\bibitem{Just2014AreMA}
Just, R., Jalali, D., Inozemtseva, L., Ernst, M.D., Holmes, R., Fraser, G.: Are
  mutants a valid substitute for real faults in software testing?
\newblock In: 22nd ACM SIGSOFT International Symposium on Foundations of
  Software Engineering, FSE 2014. Association for Computing Machinery (2014)

\bibitem{kaleeswaran:minthint:icse14}
Kaleeswaran, S., Tulsian, V., Kanade, A., Orso, A.: Minthint: Automated
  synthesis of repair hints.
\newblock In: 36th International Conference on Software Engineering, ICSE 2014,
  p. 266–276. Association for Computing Machinery (2014)

\bibitem{inproceedings-5}
Kanade, A., Maniatis, P., Balakrishnan, G., Shi, K.: Learning and evaluating
  contextual embedding of source code.
\newblock In: 37th International Conference on Machine Learning, {ICML} 2020,
  \emph{Proceedings of Machine Learning Research}, vol. 119, pp. 5110--5121.
  {PMLR} (2020)

\bibitem{Karampatsis2020HowOD}
Karampatsis, R.M., Sutton, C.: How often do single-statement bugs occur? the
  manysstubs4j dataset.
\newblock In: 17th International Conference on Mining Software Repositories,
  MSR '20, p. 573–577. Association for Computing Machinery (2020)

\bibitem{karampatsis:scelmo:20}
Karampatsis, R.M., Sutton, C.: Scelmo: Source code embeddings from language
  models.
\newblock arXiv preprint arXiv:2004.13214  (2020)

\bibitem{kingma:adam-optimizaiton:iclr15}
Kingma, D.P., Ba, J.: Adam: {A} method for stochastic optimization.
\newblock In: International Conference on Learning Representations, {ICLR}
  (2015)

\bibitem{koyuncu:ifixr:fse2019}
Koyuncu, A., Liu, K., Bissyand\'{e}, T.F., Kim, D., Monperrus, M., Klein, J.,
  Le~Traon, Y.: Ifixr: Bug report driven program repair.
\newblock In: 27th ACM Joint Meeting on European Software Engineering
  Conference and Symposium on the Foundations of Software Engineering, ESEC/FSE
  2019, p. 314–325. Association for Computing Machinery (2019)

\bibitem{legouesNFWTSE2012}
{Le Goues}, C., Nguyen, T., Forrest, S., Weimer, W.: Genprog: A generic method
  for automatic software repair.
\newblock IEEE Transactions on Software Engineering \textbf{38}, 54--72 (2012)

\bibitem{dl-identify-suspicious-return-saner-2020}
Li, G., Liu, H., Jin, J., Umer, Q.: Deep learning based identification of
  suspicious return statements.
\newblock In: 2020 IEEE 27th International Conference on Software Analysis,
  Evolution and Reengineering (SANER), pp. 480--491 (2020)

\bibitem{article-4}
Li, X., Wang, L., Xin, Y., Yang, Y., Chen, Y.: Automated vulnerability
  detection in source code using minimum intermediate representation learning.
\newblock Applied Sciences \textbf{10}, 1692 (2020)

\bibitem{li-dlfix-2020}
Li, Y., Wang, S., Nguyen, T.N.: Dlfix: Context-based code transformation
  learning for automated program repair.
\newblock In: ACM/IEEE 42nd International Conference on Software Engineering,
  ICSE '20, p. 602–614. Association for Computing Machinery (2020)

\bibitem{orange}
Lin, C.Y., Och, F.J.: Orange: a method for evaluating automatic evaluation
  metrics for machine translation.
\newblock In: The 20th International Conference on Computational Linguistics,
  pp. 501--507. COLING (2004)

\bibitem{argument-defects-icse-2016}
Liu, H., Liu, Q., Staicu, C.A., Pradel, M., Luo, Y.: Nomen est omen: Exploring
  and exploiting similarities between argument and parameter names.
\newblock In: 38th International Conference on Software Engineering, ICSE '16,
  p. 1063–1073. Association for Computing Machinery (2016)

\bibitem{Liu2019YouCF}
Liu, K., Koyuncu, A., Bissyand{\'e}, T.F., Kim, D., Klein, J., Traon, Y.L.: You
  cannot fix what you cannot find! an investigation of fault localization bias
  in benchmarking automated program repair systems.
\newblock 2019 12th IEEE Conference on Software Testing, Validation and
  Verification (ICST) pp. 102--113 (2019)

\bibitem{Liu2018LSRepairLS}
Liu, K., Koyuncu, A., Kim, K., Kim, D., Bissyand{\'e}, T.F.: Lsrepair: Live
  search of fix ingredients for automated program repair.
\newblock In: 2018 25th Asia-Pacific Software Engineering Conference (APSEC),
  pp. 658--662. IEEE Computer Society (2018)

\bibitem{liu:contextual-survey:20}
Liu, Q., Kusner, M.J., Blunsom, P.: A survey on contextual embeddings.
\newblock arXiv preprint arXiv:2003.07278  (2020)

\bibitem{article-prophet}
Long, F., Rinard, M.: Automatic patch generation by learning correct code.
\newblock In: Proceedings of the 43rd Annual ACM SIGPLAN-SIGACT Symposium on
  Principles of Programming Languages, POPL '16, p. 298–312. Association for
  Computing Machinery (2016)

\bibitem{luong17}
Luong, M., Brevdo, E., Zhao, R.: Neural machine translation (seq2seq) tutorial.
\newblock https://github.com/tensorflow/nmt  (2017)

\bibitem{luong-attention-mechanism-acl-2015}
Luong, T., Pham, H., Manning, C.D.: Effective approaches to attention-based
  neural machine translation.
\newblock In: 2015 Conference on Empirical Methods in Natural Language
  Processing, pp. 1412--1421. Association for Computational Linguistics (2015)

\bibitem{inproceedings-coconut}
Lutellier, T., Pham, H.V., Pang, L., Li, Y., Wei, M., Tan, L.: Coconut:
  Combining context-aware neural translation models using ensemble for program
  repair.
\newblock In: 29th ACM SIGSOFT International Symposium on Software Testing and
  Analysis, ISSTA 2020, p. 101–114. Association for Computing Machinery
  (2020)

\bibitem{inproceedings-CB-DS-repair}
{Malik}, M.Z., {Siddiqui}, J.H., {Khurshid}, S.: Constraint-based program
  debugging using data structure repair.
\newblock In: 2011 Fourth IEEE International Conference on Software Testing,
  Verification and Validation, pp. 190--199 (2011)

\bibitem{nl2type}
Malik, R.S., Patra, J., Pradel, M.: Nl2type: Inferring javascript function
  types from natural language information.
\newblock In: 2019 IEEE/ACM 41st International Conference on Software
  Engineering (ICSE), pp. 304--315 (2019)

\bibitem{inproceedings-sapfix}
Marginean, A., Bader, J., Chandra, S., Harman, M., Jia, Y., Mao, K., Mols, A.,
  Scott, A.: Sapfix: Automated end-to-end repair at scale.
\newblock In: 2019 IEEE/ACM 41st International Conference on Software
  Engineering: Software Engineering in Practice (ICSE-SEIP), pp. 269--278
  (2019)

\bibitem{marginean-sapFix-icse-2019}
Marginean, A., Bader, J., Chandra, S., Harman, M., Jia, Y., Mao, K., Mols, A.,
  Scott, A.: Sapfix: Automated end-to-end repair at scale.
\newblock In: 41st International Conference on Software Engineering: Software
  Engineering in Practice, ICSE-SEIP ’19, p. 269–278. IEEE Press (2019)

\bibitem{mccann:context-vector:17}
McCann, B., Bradbury, J., Xiong, C., Socher, R.: Learned in translation:
  Contextualized word vectors.
\newblock In: Advances in Neural Information Processing Systems, vol.~30.
  Curran Associates, Inc. (2017)

\bibitem{Mehne2018AcceleratingSP}
Mehne, B., Yoshida, H., Prasad, M., Sen, K., Gopinath, D., Khurshid, S.:
  Accelerating search-based program repair.
\newblock 2018 IEEE 11th International Conference on Software Testing,
  Verification and Validation (ICST) pp. 227--238 (2018)

\bibitem{mesbah-deepDelta-fse-2019}
Mesbah, A., Rice, A., Johnston, E., Glorioso, N., Aftandilian, E.: Deepdelta:
  Learning to repair compilation errors.
\newblock In: 27th ACM Joint Meeting on European Software Engineering
  Conference and Symposium on the Foundations of Software Engineering, ESEC/FSE
  2019, p. 925–936. Association for Computing Machinery (2019)

\bibitem{word2vec-iclr-2013}
Mikolov, T., Chen, K., Corrado, G., Dean, J.: Efficient estimation of word
  representations in vector space.
\newblock In: 1st International Conference on Learning Representations, {ICLR}
  2013, Workshop Track Proceedings (2013)

\bibitem{mikolov-word2vec-2013}
Mikolov, T., Sutskever, I., Chen, K., Corrado, G., Dean, J.: Distributed
  representations of words and phrases and their compositionality.
\newblock In: 26th International Conference on Neural Information Processing
  Systems - Volume 2, NIPS'13, p. 3111–3119. Curran Associates Inc. (2013)

\bibitem{monperrus:automatic-patch-generation:icse:2014}
Monperrus, M.: A critical review of "automatic patch generation learned from
  human-written patches": Essay on the problem statement and the evaluation of
  automatic software repair.
\newblock In: 36th International Conference on Software Engineering, ICSE 2014,
  p. 234–242. Association for Computing Machinery (2014)

\bibitem{article-bibliography}
Monperrus, M.: Automatic software repair: A bibliography.
\newblock ACM Comput. Surv. \textbf{51}(1) (2018)

\bibitem{Nguyen2015GraphBasedSL}
Nguyen, A.T., Nguyen, T.N.: Graph-based statistical language model for code.
\newblock In: 37th International Conference on Software Engineering - Volume 1,
  ICSE '15, p. 858–868. IEEE Press (2015)

\bibitem{tree-lstm:iclr:2020}
Nguyen, X., Joty, S.R., Hoi, S.C.H., Socher, R.: Tree-structured attention with
  hierarchical accumulation.
\newblock In: 8th International Conference on Learning Representations, {ICLR}
  2020. OpenReview.net (2020)

\bibitem{articlepan}
Pan, K., Kim, S., Whitehead, E.J.: Toward an understanding of bug fix patterns.
\newblock Empirical Software Engineering \textbf{14}(3), 286–315 (2009)

\bibitem{pandita-name-based-inferring-constraints-from-naming-2016}
Pandita, R., Taneja, K., Williams, L., Tung, T.: Icon: Inferring temporal
  constraints from natural language api descriptions.
\newblock In: 2016 IEEE International Conference on Software Maintenance and
  Evolution (ICSME), pp. 378--388 (2016)

\bibitem{pandita-name-based-method-specification-from-naming-2012}
Pandita, R., Xiao, X., Zhong, H., Xie, T., Oney, S., Paradkar, A.: Inferring
  method specifications from natural language api descriptions.
\newblock In: 2012 34th International Conference on Software Engineering
  (ICSE), pp. 815--825 (2012)

\bibitem{Papineni2002BleuAM}
Papineni, K., Roukos, S., Ward, T., Zhu, W.J.: Bleu: A method for automatic
  evaluation of machine translation.
\newblock In: 40th Annual Meeting on Association for Computational Linguistics,
  ACL '02, p. 311–318. Association for Computational Linguistics (2002)

\bibitem{inproceedings-glove}
Pennington, J., Socher, R., Manning, C.: {G}lo{V}e: Global vectors for word
  representation.
\newblock In: 2014 Conference on Empirical Methods in Natural Language
  Processing ({EMNLP}), pp. 1532--1543. Association for Computational
  Linguistics (2014)

\bibitem{article-elmo}
Peters, M.E., Neumann, M., Iyyer, M., Gardner, M., Clark, C., Lee, K.,
  Zettlemoyer, L.: Deep contextualized word representations.
\newblock In: 2018 Conference of the North {A}merican Chapter of the
  Association for Computational Linguistics: Human Language Technologies,
  Volume 1 (Long Papers), pp. 2227--2237. Association for Computational
  Linguistics (2018)

\bibitem{peters:elmo:18}
Peters, M.E., Neumann, M., Iyyer, M., Gardner, M., Clark, C., Lee, K.,
  Zettlemoyer, L.: Deep contextualized word representations.
\newblock In: 2018 Conference of the North {A}merican Chapter of the
  Association for Computational Linguistics: Human Language Technologies,
  Volume 1 (Long Papers), pp. 2227--2237. Association for Computational
  Linguistics (2018)

\bibitem{pewny-evilcoder-2016}
Pewny, J., Holz, T.: Evilcoder: Automated bug insertion.
\newblock In: Proceedings of the 32nd Annual Conference on Computer Security
  Applications, ACSAC '16, p. 214–225. Association for Computing Machinery
  (2016)

\bibitem{argument-defects-issta-2011}
Pradel, M., Gross, T.R.: Detecting anomalies in the order of equally-typed
  method arguments.
\newblock In: 2011 International Symposium on Software Testing and Analysis,
  ISSTA '11, p. 232–242. Association for Computing Machinery (2011)

\bibitem{arguments-defects-tse-2013}
Pradel, M., Gross, T.R.: Name-based analysis of equally typed method arguments.
\newblock IEEE Transactions on Software Engineering \textbf{39}(8), 1127--1143
  (2013)

\bibitem{pradel-deepbugs-oopsla-2018}
Pradel, M., Sen, K.: Deepbugs: A learning approach to name-based bug detection.
\newblock Proc. ACM Program. Lang. \textbf{2}(OOPSLA) (2018)

\bibitem{veselin-learning-programs-from-noisy-data-2016}
Raychev, V., Bielik, P., Vechev, M., Krause, A.: Learning programs from noisy
  data.
\newblock SIGPLAN Not. \textbf{51}(1), 761–774 (2016)

\bibitem{Predictingprogrampropertiesfrombigcode}
Raychev, V., Vechev, M., Krause, A.: Predicting program properties from "big
  code".
\newblock In: 42nd Annual ACM SIGPLAN-SIGACT Symposium on Principles of
  Programming Languages, POPL '15, p. 111–124. Association for Computing
  Machinery (2015)

\bibitem{inproceedings-bert}
Reimers, N., Gurevych, I.: Sentence-{BERT}: Sentence embeddings using {S}iamese
  {BERT}-networks.
\newblock In: 2019 Conference on Empirical Methods in Natural Language
  Processing and the 9th International Joint Conference on Natural Language
  Processing (EMNLP-IJCNLP), pp. 3982--3992. Association for Computational
  Linguistics (2019)

\bibitem{argument-selection-defects-oopsla-2017}
Rice, A., Aftandilian, E., Jaspan, C., Johnston, E., Pradel, M.,
  Arroyo-Paredes, Y.: Detecting argument selection defects.
\newblock Proc. ACM Program. Lang. \textbf{1}(OOPSLA) (2017)

\bibitem{article-AR-PHP}
Samimi, H., Sch\"{a}fer, M., Artzi, S., Millstein, T., Tip, F., Hendren, L.:
  Automated repair of html generation errors in php applications using string
  constraint solving.
\newblock In: 34th International Conference on Software Engineering, ICSE '12,
  p. 277–287. IEEE Press (2012)

\bibitem{sato:expressiveness-gnn:20}
Sato, R.: A survey on the expressive power of graph neural networks.
\newblock arXiv preprint arXiv:2003.04078  (2020)

\bibitem{inproceedings-learned-heuristics}
Schramm, L.: Improving performance of automatic program repair using learned
  heuristics.
\newblock In: 11th Joint Meeting on Foundations of Software Engineering,
  ESEC/FSE 2017, p. 1071–1073. Association for Computing Machinery (2017)

\bibitem{detecting-swapped-arguments-scam-2020}
Scott, R., Ranieri, J., Kot, L., Kashyap, V.: Out of sight, out of place:
  Detecting and assessing swapped arguments.
\newblock In: 2020 IEEE 20th International Working Conference on Source Code
  Analysis and Manipulation (SCAM), pp. 227--237 (2020)

\bibitem{sellik:off-by-one-msr:21}
Sellik, H., van Paridon, O., Gousios, G., Aniche, M.: Learning off-by-one
  mistakes: An empirical study.
\newblock In: 2021 IEEE/ACM 18th International Conference on Mining Software
  Repositories (MSR), pp. 58--67 (2021)

\bibitem{vulnerability:tang:2020}
Tang, G., Meng, L., Wang, H., Ren, S., Wang, Q., Yang, L., Cao, W.: A
  comparative study of neural network techniques for automatic software
  vulnerability detection.
\newblock In: 2020 International Symposium on Theoretical Aspects of Software
  Engineering (TASE), pp. 1--8. IEEE Computer Society (2020)

\bibitem{tarlow-graph-2-diff-2019}
Tarlow, D., Moitra, S., Rice, A., Chen, Z., Manzagol, P.A., Sutton, C.,
  Aftandilian, E.: Learning to fix build errors with graph2diff neural
  networks.
\newblock In: Proceedings of the IEEE/ACM 42nd International Conference on
  Software Engineering Workshops, p. 19–20. Association for Computing
  Machinery (2020)

\bibitem{Learning-Meaningful-Code-Changes-Via-NMT-2019-7}
Tufano, M., Pantiuchina, J., Watson, C., Bavota, G., Poshyvanyk, D.: On
  learning meaningful code changes via neural machine translation.
\newblock In: 41st International Conference on Software Engineering, ICSE '19,
  p. 25–36. IEEE Press (2019)

\bibitem{inproceedings-6}
Tufano, M., Watson, C., Bavota, G., Di~Penta, M., White, M., Poshyvanyk, D.:
  Deep learning similarities from different representations of source code.
\newblock In: 2018 IEEE/ACM 15th International Conference on Mining Software
  Repositories (MSR), pp. 542--553 (2018)

\bibitem{inproceedings-wild}
Tufano, M., Watson, C., Bavota, G., Di~Penta, M., White, M., Poshyvanyk, D.: An
  empirical investigation into learning bug-fixing patches in the wild via
  neural machine translation.
\newblock In: 33rd ACM/IEEE International Conference on Automated Software
  Engineering, ASE 2018, p. 832–837. Association for Computing Machinery
  (2018)

\bibitem{tufano:tosem:19}
Tufano, M., Watson, C., Bavota, G., Penta, M.D., White, M., Poshyvanyk, D.: An
  empirical study on learning bug-fixing patches in the wild via neural machine
  translation.
\newblock ACM Transactions on Software Engineering and Methodology  (2019)

\bibitem{code-review-icse-2021}
Tufano, R., Pascarella, L., Tufano, M., Poshyvanyk, D., Bavota, G.: Towards
  automating code review activities.
\newblock In: 2021 IEEE/ACM 43rd International Conference on Software
  Engineering (ICSE), pp. 163--174. IEEE Computer Society (2021)

\bibitem{article-jointly-learning}
Vasic, M., Kanade, A., Maniatis, P., Bieber, D., singh, R.: Neural program
  repair by jointly learning to localize and repair.
\newblock In: International Conference on Learning Representations.
  OpenReview.net (2019)

\bibitem{transformer:neurips:2017}
Vaswani, A., Shazeer, N., Parmar, N., Uszkoreit, J., Jones, L., Gomez, A.N.,
  Kaiser, L.u., Polosukhin, I.: Attention is all you need.
\newblock In: Advances in Neural Information Processing Systems, vol.~30.
  Curran Associates, Inc. (2017)

\bibitem{velickovic:gat:18}
Veličković, P., Cucurull, G., Casanova, A., Romero, A., Liò, P., Bengio, Y.:
  Graph attention networks.
\newblock In: International Conference on Learning Representations (2018)

\bibitem{wainakh-semantic-representations-of-source-code-2019}
Wainakh, Y., Rauf, M., Pradel, M.: Idbench: Evaluating semantic representations
  of identifier names in source code.
\newblock In: 2021 IEEE/ACM 43rd International Conference on Software
  Engineering (ICSE), pp. 562--573 (2021)

\bibitem{wang-coset-2019}
Wang, K., Christodorescu, M.: {COSET:} {A} benchmark for evaluating neural
  program embeddings.
\newblock arXiv preprint arXiv:1905.11445  (2019)

\bibitem{watson-learning-meaningful-assert-statements-icse-2020}
Watson, C., Tufano, M., Moran, K., Bavota, G., Poshyvanyk, D.: On learning
  meaningful assert statements for unit test cases.
\newblock In: ACM/IEEE 42nd International Conference on Software Engineering,
  ICSE '20, p. 1398–1409. Association for Computing Machinery (2020)

\bibitem{article-white}
{White}, M., {Tufano}, M., {Martínez}, M., {Monperrus}, M., {Poshyvanyk}, D.:
  Sorting and transforming program repair ingredients via deep learning code
  similarities.
\newblock In: 2019 IEEE 26th International Conference on Software Analysis,
  Evolution and Reengineering (SANER), pp. 479--490 (2019)

\bibitem{wohlin2012experimentation}
Wohlin, C., Runeson, P., H{\"o}st, M., Ohlsson, M.C., Regnell, B., Wessl{\'e}n,
  A.: Experimentation in software engineering.
\newblock Springer Science \& Business Media (2012)

\bibitem{article-constraint-model}
Wotawa, F., Nica, M., Nica, I.: Automated debugging based on a constraint model
  of the program and a test case.
\newblock The Journal of Logic and Algebraic Programming \textbf{81}, 390–407
  (2012)

\bibitem{gnn-survey:2021}
Wu, Z., Pan, S., Chen, F., Long, G., Zhang, C., Yu, P.S.: A comprehensive
  survey on graph neural networks.
\newblock IEEE Transactions on Neural Networks and Learning Systems
  \textbf{32}(1), 4--24 (2021)

\bibitem{subtree}
Zhang, J., Wang, X., Zhang, H., Sun, H., Wang, K., Liu, X.: A novel neural
  source code representation based on abstract syntax tree.
\newblock In: ICSE, pp. 783--794 (2019)

\bibitem{zhao:neural-edits:icrl:19}
Zhao, R., Bieber, D., Swersky, K., Tarlow, D.: Neural networks for modeling
  source code edits.
\newblock arXiv preprint arXiv:1904.02818  (2019)

\bibitem{zhou:gnnsurvey}
Zhou, J., Cui, G., Hu, S., Zhang, Z., Yang, C., Liu, Z., Wang, L., Li, C., Sun,
  M.: Graph neural networks: A review of methods and applications.
\newblock AI Open \textbf{1}, 57--81 (2020)

\end{thebibliography}

\parpic{\includegraphics[width=1in,clip,keepaspectratio]{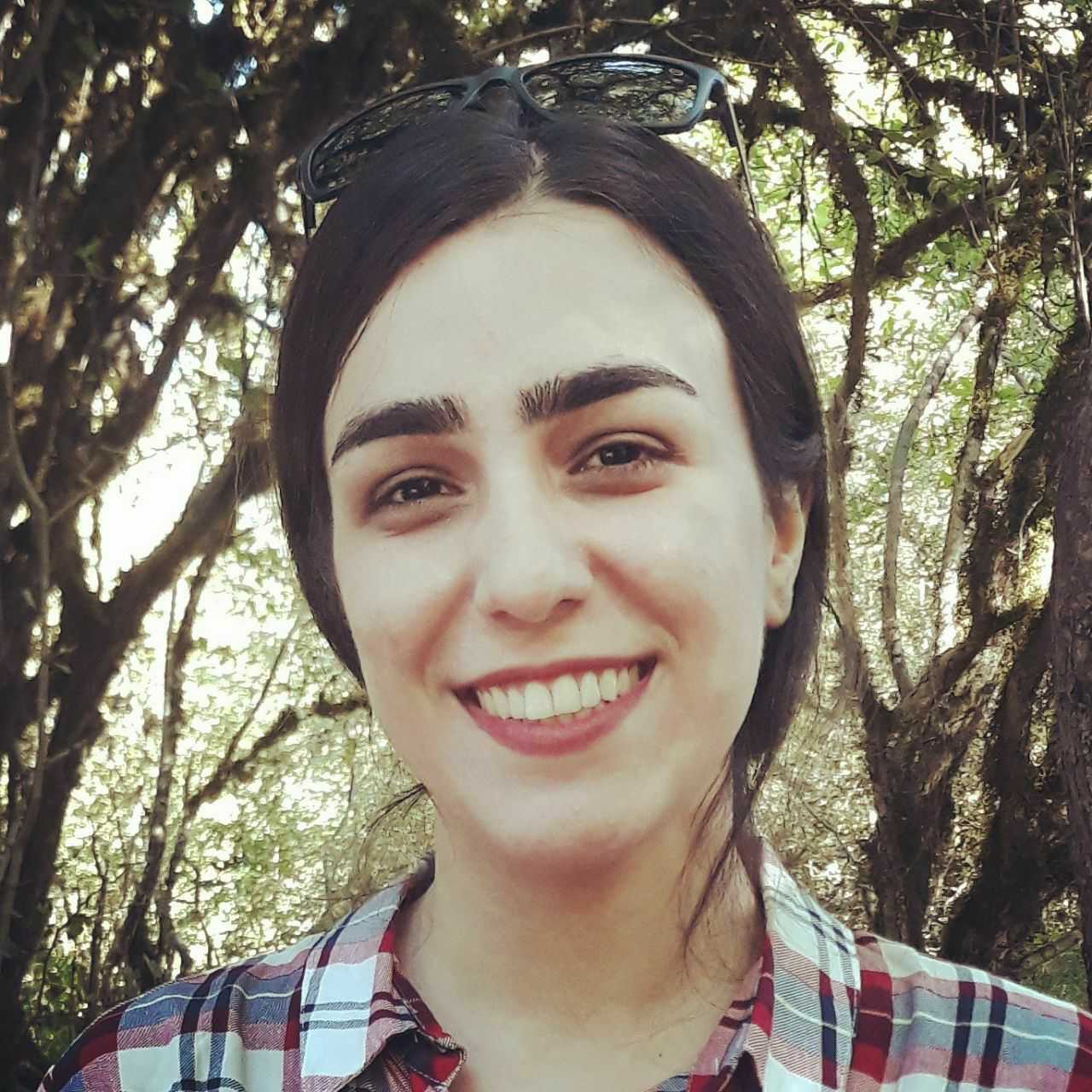}}
\noindent {\bf Marjane Namavar} is a software engineer. She received her M.A.Sc. degree in 2021 from the School of Electrical and Computer Engineering, University of British Columbia, Canada, under the supervision of Professor Ali Mesbah. Her research interests include the applications of machine learning in source code analysis and repair.

\vspace{2\baselineskip}
\parpic{\includegraphics[width=1in,clip,keepaspectratio]{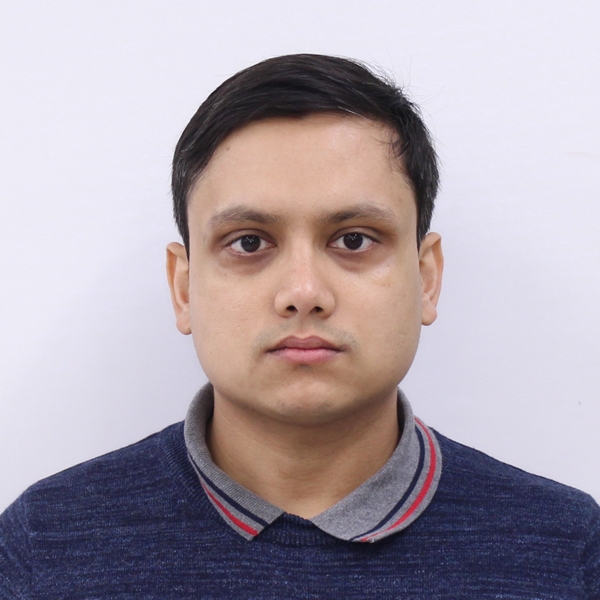}}
\noindent {\bf Noor Nashid} Noor Nashid is a Ph.D. student at the University of British Columbia (UBC). His research interests include software engineering, with an emphasis on the applications of deep learning techniques to source code. He is a recipient of the four-year doctoral fellowship at UBC.

\vspace{2\baselineskip}
\parpic{\includegraphics[width=1in,clip,keepaspectratio]{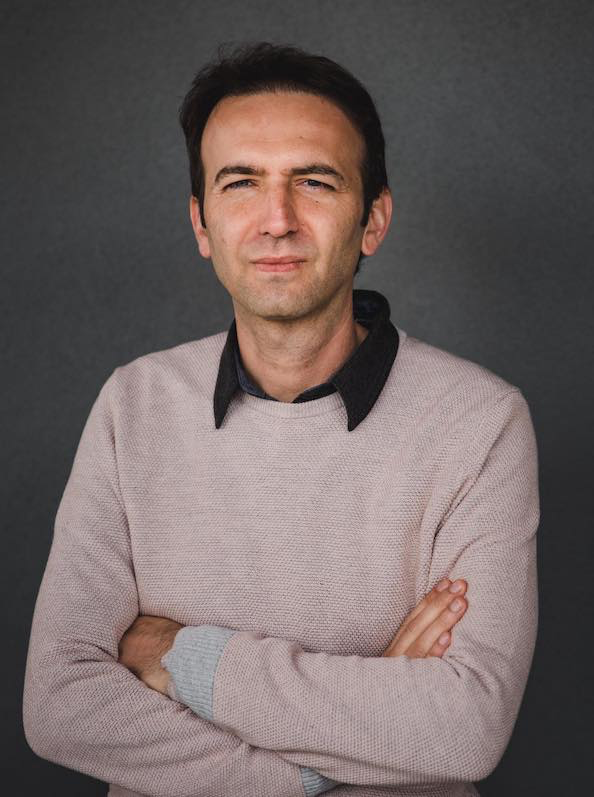}}
\noindent {\bf Ali Mesbah} is a professor at the University of British Columbia (UBC) where he leads the Software Analysis and Testing (SALT)
research lab. His main area of research is in software engineering and his research interests include software analysis and testing, web and mobile-based applications, software maintenance and evolution, debugging and fault localization, and automated program repair. He has published over 60 peer-reviewed papers and received numerous best paper awards, including two ACM Distinguished Paper Awards at the International Conference on Software Engineering (ICSE 2009 and ICSE 2014). He is the recipient of a Killam Accelerator Research Fellowship (KARF) award (2020), a Killam Faculty Research Prize (2019) at UBC, and was awarded the NSERC Discovery Accelerator Supplement (DAS) award in 2016. He is currently an associate editor of the IEEE Transactions on Software Engineering (TSE) and regularly serves on the program committees of numerous software engineering conferences such as ICSE, FSE, ASE, ISSTA, and ICST.

\end{document}